\documentclass[%
reprint,
superscriptaddress,
 amsmath,amssymb,
 aps,
prd,
]{revtex4-2}

\usepackage[utf8]{inputenc}
\usepackage{graphicx}
\usepackage{dcolumn}
\usepackage{multirow}
\usepackage{bm}
\usepackage{mathrsfs}
\usepackage{hyperref}
\hypersetup{
  colorlinks = true,
  urlcolor = blue,
  linkcolor = blue,
  citecolor = cyan,
  filecolor = magenta,
}

\newcommand{\ud}{\mathrm{d}}
\newcommand{\pd}{\partial}

\newcommand{\lie}{\mathscr{L}}

\newcommand{\order}[1]{\mathcal{O}\left(#1\right)}

\newcommand{\sD}{\mathscr{D}}

\newcommand{\sP}{\mathscr{P}}
\newcommand{\hsP}{\hat{\mathscr{P}}}

\newcommand{\fL}{\mathfrak{L}}

\begin{document}


\title{Finitely supertranslated Schwarzschild black hole and its perturbations}

\author{Shaoqi Hou}
\email{hou.shaoqi@whu.edu.cn}
\affiliation{School of Physics and Technology, Wuhan University, Wuhan, Hubei 430072, China}
\author{Kai Lin}
\email{lk314159@hotmail.com}
\affiliation{Universidade Federal de Campina Grande, Campina Grande, PB, Brasil}
\author{Zong-Hong Zhu}
\email{zhuzh@whu.edu.cn}
\affiliation{School of Physics and Technology, Wuhan University, Wuhan, Hubei 430072, China}
\affiliation{Department of Astronomy, Beijing Normal University, Beijing 100875,  China}

\date{\today}

\begin{abstract}
  A finitely supertranslated Schwarzschild black hole possesses nontrivial super-Lorentz charges compared with the standard one.
  This may impact the quasinormal modes of the black hole.
  Since the Einstein's equations are generally covariant, the quasinormal modes of a supertranslated black hole can be obtained by supertranslating the familiar results for a standard black hole.
  It turns out that the supertranslated quasinormal modes can be obtained by simply shifting the retarded time of the standard modes by an angle-dependent function parameterizing the supertranslation.
  Therefore, the supertranslated quasinormal modes oscillate at the same frequencies and decay at the same rates as the corresponding standard ones.
  The supertranslated metric is time translation invariant, but does not explicitly respect spherical symmetries, although it is implicitly rotationally symmetric.
  So the supertranslated perturbations can still be written as linear combinations of the eigenfunctions of the generalized angular momentum operators for the underlying rotational symmetry.
  With a suitably  defined asymptotic parity transformation, any perturbation can be decomposed into the even and odd parity parts.
  Then, one may conclude that the isospectrality still holds.
  To detect such supertranslated quasinormal modes, one has to place multiple gravitational wave interferometers around the supertranslated black hole, and measure the differences in the time shifts between interferometers.
  Gravitational lensing may also be helpful in the same spirit.
\end{abstract}

\maketitle


\section{Introduction}

The no-hair theorem states that all isolated, stationary black hole spacetimes are diffeomorphic to Kerr spacetimes \cite{Chrusciel:2012jk}.
So it has long been believed that such a black hole is characterized solely by its mass and angular momentum, and the diffeomorphisms that transform these black hole solutions are pure gauge.
However, the more recent studies of the infrared structure of the gravitating system have revealed a different viewpoint \cite{Strominger:2014pwa,Strominger2014bms,He:2014laa,Strominger:2018inf}.
The Bondi-Metzner-Sachs (BMS) coordinate transformations \cite{Bondi:1962px,Sachs:1962wk,Sachs1962asgr} act on the black hole spacetime nontrivially.

As an asymptotically flat spacetime, an isolated gravitating system possesses the asymptotic symmetries, named the BMS symmetries, which include supertranslations and Lorentz transformations \cite{Penrose:1962ij,Bondi:1962px,Sachs:1962wk,Penrose:1965am,Geroch1977,Wald:1984rg}.
Supertranslations can be viewed as the angle-dependent translations, and the Lorentz transformations conformally transform the metric on a unit 2-sphere \cite{Sachs1962asgr,Barnich:2010eb}.
The BMS group is the semidirect product of the Lorentz group by the supertranslation group.
Computed using suitable covariant phase method \cite{Ashtekar:1981bq,Wald:1999wa,Barnich:2011mi,Flanagan:2015pxa,Alessio:2019cae}, the conserved charges associated with the BMS symmetries depend nontrivially on the isolated systems, and vary from one isolated system to another.
In particular, isolated systems without gravitational waves  are physically different from each other, if they carry different charges.
Due to the discovery of the spin and center-of-mass memories \cite{Pasterski:2015tva,Nichols:2018qac}, extended BMS group \cite{Barnich:2009se,Barnich:2010eb,Barnich:2011ct} and generalized BMS group \cite{Campiglia:2014yka,Campiglia:2015yka,Campiglia:2020qvc} were proposed.
The Lorentz group is replaced with the Virasoro group or the diffeomorphism group of the 2-sphere.
Like the supertranslation, let us call  the super-Lorentz transformations the elements in the Virasoro group or the diffeomorphism group of $S^2$.
The transformations in these groups will be simply called the asymptotic symmetries, and the charges  the asymptotic charges.
The Weyl-BMS group is even larger, encompassing the previous ones \cite{Freidel:2021fxf}.
But in this work, let us focus on the smaller ones.

An asymptotically flat spacetime usually is determined by quantities such as the Bondi mass aspect $\mathcal M$, the shear tensor $c_{AB}$ and the angular momentum aspect $\mathcal N_A$ in the Bondi-Sachs coordinates $(u,r,\theta,\phi)$, where $A,B=\theta,\phi$ \cite{Bondi:1962px,Sachs:1962wk}.
They are generally functions of the retarded time $u$ and the angles $(\theta,\phi)$.
In a stationary spacetime without matter, it is possible to find a Bondi frame such that $\mathcal M$ is a constant, $\mathcal N_A$ is $u$-independent, and importantly, $c_{AB}$ is of the electric parity type \cite{Flanagan:2015pxa}.
Such a triplet $(\mathcal M,c_{AB},\mathcal N_A)$ describes a gravitational vacuum.
As mentioned in the previous paragraph, there can be physically distinct gravitational vacua, as long as their asymptotic charges are different.

These gravitational vacua may transition to each other.
This was uncovered due to the study of gravitational memory effect, which describes the permanent change in the arm lengths of the interferometer.
Although this effect was identified theoretically a long time ago \cite{Zeldovich:1974gvh,Kovacs:1978eu,Braginsky:1986ia,1987Natur.327..123B,Christodoulou1991,Wiseman:1991ss,Blanchet:1992br,Thorne:1992sdb}, only until the past few years, it is viewed as the vacuum transition, measured by a suitable supertranslation transformation \cite{Strominger:2014pwa,Strominger2014bms,He:2014laa,Strominger:2018inf}.
The magnitude of the memory effect is determined by the null energy flux, conjugate to the supertranslation.
The memory effect indicates the presence of the degenerate gravitational vacua at the null infinity \cite{Strominger:2014pwa}.
One can supertranslate a vacuum, and a physically distinguishable vacuum is produced.
There are infinitely many vacua.

The search for memory effects has been conducted, but no strong evidence has been found by LIGO-Virgo-KAGRA collaboration \cite{Lasky:2016knh,Yang:2018ceq,Hubner:2019sly,Zhao:2021hmx,Islam:2021old,Hubner:2021amk,Cheung:2024zow} or the North American Nanohertz Observatory for Gravitational Waves (NANOGrav) \cite{NANOGrav:2023vfo}.
The prospects of its future detection have been anticipated by several works.
It was found out that with the networks of the ground-based interferometers, one may observe the memory effect by coherently adding the signals from hundreds or thousands stellar mass binary black hole merger events after 3 - 5 years operation \cite{Johnson:2018xly,Hubner:2019sly,Grant:2022bla}.
TianQin \cite{Luo:2015ght,Sun:2022pvh} and LISA \cite{Gasparotto:2023fcg,Inchauspe:2024ibs} are able to detect the memory signals produced by massive binary black hole mergers.
Notably, it was recently predicted that DECIGO could detect a few thousand strong enough memory signals produced by stellar-mass binary black holes \cite{Hou:2024rgo}.

The detection of the memory effect would provide the observational evidence of the presence of the gravitational vacuum transition, but the physical consequences of these vacua may not be fully revealed in the memory effect.
So in this work, we will be interested in the impact of the vacua carrying nontrivial asymptotic charges.
The focus will be on stationary black hole spacetimes in general relativity, i.e., Schwarzschild black holes.
In the context of the black hole physics, the asymptotic charges are often referred as the soft hairs \cite{Hawking:2016msc}.
Since the supertranslated Schwarzschild black hole will be considered, the familiar metric found in many textbooks, e.g. \cite{Wald:1984rg,Weinberg:1972kfs,Carroll:2004st}, is said to be standard.
We will perturb the supertranslated Schwarzschild black hole, and discuss the properties of the quasinormal modes.
We would also propose two possible methods to measure the perturbations to infer the presence of the supertranslated black hole.

In fact, the physical influence of a supertranslated black hole was already considered previously.
For example, Hawking, Perry and Strominger \cite{Hawking:2016msc} argued the soft hair of a supertranslated Vaidya black hole \cite{Vaidya:1951zza} may resolve the information paradox \cite{Hawking:1976ra}.
The Hawking radiation of this black hole was later derived with the tunneling method in Ref.~\cite{Chu:2018tzu}, and its spectrum depends on the soft hair.
The Hawking radiation of a supertranslated Schwarzschild black hole was also studied in Ref.~\cite{Iofa:2017ukq}, and the spectrum is not altered.
In these studies, the infinitesimally supertranslated black holes were considered.
In the current work, the focus will be on a finitely supertranslated Schwarzschild black hole.
One shall note that an infinitesimally supertranslated metric differs from the original one by a small perturbation, while a finitely supertranslated one differs by an arbitrarily large amount.
Since the gravitation is nonlinear, it is interesting to study the perturbations of a finitely supertranslated black hole.

In Ref.~\cite{Compere:2016hzt}, such a metric was given in a closed form,
and some of its kinematic properties were scrutinized.
In a suitably defined coordinate system $(t,\rho,z^A)$, related to the Schwarzschild coordinates via Eq.~(26) in Ref.~\cite{Compere:2016hzt}, the metric is
\begin{subequations}
  \label{eq-st-sch-s}
  \begin{gather}
    \ud\mathring s^2=-\left(\frac{1-M/2\rho_s}{1+M/2\rho_s}\right)^2\ud t^2+\left(1+\frac{M}{2\rho_s}\right)^4\ud\bm l^2,\\
    \ud\bm l^2=\ud \rho^2+\{[(\rho-E)^2+U]\gamma_{AB}+(\rho-E)c_{AB}\}\ud z^A\ud z^B,
  \end{gather}
\end{subequations}
where $M$ is the Schwarzschild mass, $\gamma_{AB}$ is the round metric on a unit 2-sphere with $\sD_A$ its derivative, $\beta=\beta(z^A)$ is arbitrary, $\beta_{(0,0)}$ is the lowest spherical mode of $\beta$, $\rho_s=[(\rho-\beta+\beta_{(0,0)})^2+\sD_A\beta\sD^A\beta]^{1/2}$, $c_{AB}=-(2\sD_A\sD_B-\gamma_{AB}\sD_C\sD^C)\beta$, $U=c_A^Bc_B^A/8$, and $E=\sD_A\sD^A\beta/2+\beta-\beta_{(0,0)}$.
One may directly work out the perturbations to this metric in this coordinate system.
Unfortunately, this metric does not explicitly possess any symmetry, except the time translation symmetry.
It would be tremendously difficult to solve the perturbed Einstein's equation.
Even if one can successfully reduce the perturbed Einstein's equation, one immediately comes across another difficulty.
That is, in addition to the event horizon, there are also new coordinate singularities.
These singularities form the supertranslation horizon, which generally differs from and intersects with the event horizon.
This implies the difficulty in imposing the boundary condition at the event horizon.
So in this work, this metric will not be used.

Instead, one shall take advantage of the general covariance of the Einstein's equation.
That is, if a solution to Einstein's equation is determined in one coordinate system, the transformed solution under a diffeomorphism solves the transformed Einstein's equation.
Therefore, the strategy here is to take the standard perturbative solutions of the Schwarzschild spacetime and transform them to the suitable Bondi-Sachs coordinates, in which the background metric describes the supertranslated black hole.
Then, one obtains the supertranslated perturbations.
One cannot simply apply the coordinate transformation provided by Ref.~\cite{Compere:2016hzt}, as it generally does not transform the perturbed Schwarzschild metric to the desired form.
And even worse, one cannot easily generalize the method in Ref.~\cite{Compere:2016hzt} to get a new coordinate transformation for the perturbed metric.
The very coordinate transformation derived in that work heavily relies  on the fact that the background metric is time-independent, while the perturbed metric is usually not.

Since we will be eventually interested in detecting the presence of a supertranslated black hole by observing its perturbations, we merely have to perform a suitable coordinate transformation near the null infinity, where the measurement is conducted.
Such a finite supertranslation has already been derived in Ref.~\cite{Flanagan:2023jio} for a generic asymptotically flat spacetime.
Applying this finite supertranslation transformation would give the perturbations for the supertranslated Schwarzschild black hole.
It was thus found out that the supertranslated perturbation can be easily obtained by shifting the retarded time $u$ of the standard results by a function of angles, $\beta(\theta,\phi)$, which quantifies the supertranslation.
Like Eq.~\eqref{eq-st-sch-s}, the supertranslated metric obtained with a finite supertranslation is not invariant under the parity transformation.
In fact, it might not be sensible to have some transformation named parity.
Nevertheless, by studying the property of the parity transformation in the usual Schwarzschild spacetime, we proposed a new transformation named asymptotic parity transformation.
Under this transformation, it is meaningful to state that the spectra of the odd and even parity perturbations are the same as the standard ones, and the isospectrality is not violated.

Note that we perform the finite supertranslation transformation in the active manner.
This means that one treats the coordinate transformation in terms of a diffeomorphism $\varphi$ from the manifold $\mathscr M$ to itself.
Its pullback $\varphi_*$ maps the metric $g$ to $\varphi_*g$ \cite{Wald:1984rg}.
$\varphi_*g$ is viewed as a new metric on $\mathscr M$.
The two pairs $(\mathscr M,g)$ and $(\mathscr M,\varphi_*g)$ are two spacetimes \cite{Wald:1984rg}.
As they share the same manifold $\mathscr M$, both $g$ and $\varphi_*g$ can be expressed using the same coordinate chart at the same point in $\mathscr M$.
This is also true for other tensor fields, such as the Killing vector fields $\chi$ of the standard Schwarzschild spacetime.
In contrast, a passive coordinate transformation is to relabel a spacetime point  by a different set of numbers (coordinates), so the metric $g$ and any other tensors remain the same, except that their components in the new coordinates change.
Of course, mathematically, the active and passive transformations are interconnected together in a diffeomorphism invariant theory.
This allows us to perform the active supertranslation just as if we passively transform the original metric.
This mathematical manipulation would lead to physically different spacetimes, as long as the transformation involves the BMS transformation or its generalizations.

This work is organized in the following way.
After the presentation of the notations and conventions used in this work in Sec.~\ref{sec-cn}, some basics of the asymptotically flat spacetime will be reviewed in Sec.~\ref{sec-bs}.
The finite supertranslation will be discussed in Sec.~\ref{sec-f-st}.
Then, Sec.~\ref{sec-sch} will be devoted to the discussion of the standard Schwarzschild metric (Sec.~\ref{sec-sd-sch}) and the supertranslated one (Sec.~\ref{sec-st-sbh}).
Section~\ref{sec-bhper} reviewed the black hole perturbation theory, in particular, the gauge-invariant formalism.
After that, the standard black hole perturbations are expressed in the suitable Bondi-Sachs coordinates in Sec.~\ref{sec-tbs}.
Since there are two types of perturbations, we start with the even parity perturbation in Sec.~\ref{sec-e-bs}, and then the odd parity perturbation in Sec.~\ref{sec-o-bs}.
Section~\ref{sec-sum} is a short intermediate summary.
The supertranslated perturbations are obtained in Sec.~\ref{sec-stper}, and their spherical decomposition is discussed in Sec.~\ref{sec-sph-dec}.
In Section~\ref{sec-det1}, we speculate the possible ways to detect the supertranslated perturbation.
Finally, there is a conclusion in \ref{sec-con}.
In Appendix~\ref{sec-aop}, the generalized angular momentum operators are defined for tensors on the unit 2-sphere.
In Appendix~\ref{app-bdyc}, the covariant boundary conditions for the quasinormal modes are discussed.

\subsection{Notations and conventions}
\label{sec-cn}

In the following sections, multiple coordinate systems will be used, including the usual Schwarzschild coordinates $(t,r,\theta,\phi)$, the retarded coordinates $(u,r,\theta,\phi)$ and the advanced coordinates $(v,r,\theta,\phi)$.
For the standard Schwarzschild metric, the retarded and the advanced coordinates are actually the Bondi-Sachs coordinates.
For notational simplicity, we use $\theta^A$ to collectively represent $(\theta,\phi)$, and the capital Latin indices $A,B,\cdots=1,2$ to label the angular directions.
Often, $\bm\theta=(\theta,\phi)$ is used.
Following Ref.~\cite{Martel:2005ir}, $x^a=(t,r)$, or $x^a=(u,r)$, or $x^a=(v,r)$, depending on the context.
After a BMS transformation, the retarded Bondi-Sachs coordinates are $(\hat u,\hat r,\hat\theta^A)$.

In the unhatted coordinates, the metric is denoted by $g_{\mu\nu}$ and its line element by $\ud s^2$, while in the hatted coordinates, they are $\hat g_{\mu\nu}$ and $\ud \hat s^2$, respectively.
For the background metric, its line element is $\ud\mathring s^2$ in the unhatted coordinates and $\ud\hat{\mathring s}^2$ in the hatted ones.
Like the line element, many quantities will be hatted if they are expressed in the hatted coordinates.
The round metric on a unit 2-sphere is $\gamma_{AB}$, which is $\gamma_{AB}\ud\theta^A\ud\theta^B=\ud\theta^2+\sin^2\theta\ud\phi^2$.
Its inverse is $\gamma^{AB}$, volume element $\epsilon_{AB}$, determinant $\gamma$, and covariant derivative $\sD_A$.
The capital Latin indices will be raised or lowered with $\gamma^{AB}$ or $\gamma_{AB}$, respectively.
For the standard Schwarzschild metric, its temporal-radial components are labeled by $\bar g_{ab}$, whose inverse is $\bar g^{ab}$, volume element $\bar\epsilon_{ab}$ and covariant derivative $\bar\nabla_a$.
The lower case Latin indices will be naturally raised or lowered with $\bar g^{ab}$ or $\bar g_{ab}$, respectively.

Usually, the generator of the super-Lorentz transformation is donated by $Y^A$.
Since in the black hole perturbation theory, spherical harmonics $Y^{\ell m}$ will be frequently used, we instead use $\mathcal Y^A$ to represent a super-Lorentz generator.
We also use $\mathcal M$ to denote a generic Bondi mass aspect, and reserve $M$ for the Schwarzschild mass.

An overhead dot implies $\pd_u$, and the double dot means $\pd_u^2$.
$[\cdot]$ totally antisymmetrizes the indices, and $(\cdot)$ symmetrizes the indices.
$\langle\cdot\rangle$ enclosing the capital Latin indices is to take the symmetric, traceless part.

Finally, within the Bondi-Sachs formalism, one always expands the components of the metric and other tensor fields in $1/r$, and retains the leading order terms.
So we will use $\cdots$ to represent higher order terms $1/r$.
Since we will always work up to the linear order in the metric perturbation, higher order contributions in the metric perturbation will never be presented in any manner.

We use the natural units with $G=c=1$.

\section{Asymptotically flat spacetimes}
\label{sec-bs}

Roughly speaking, an asymptotically flat spacetime is the one approaching the Minkowski spacetime at the large distances from the source of gravity.
There are five types of infinities in Minkowski spacetime, the spatial infinity ($i^0$), the future and past timelike infinities ($i^+$ and $i^-$), and the future and past null inifnities ($\mathscr I^+$ and $\mathscr I^-$) \cite{Wald:1984rg}.
The asymptotic flatness is usually defined at the spatial and null infinities.
For the purpose of studying the black hole quasinormal modes, the focus will be on the future null infinity $\mathscr I^+$, and the Bondi-Sachs formalism will be used.
For more formal definitions of asymptotic flatness, please refer to Refs.~\cite{Penrose:1962ij,Penrose:1965am,Penrose:1985bww,Geroch1977,Ashtekar:1978zz,Wald:1984rg} for instance.

In Bondi-Sachs formalism, the metric is written as \cite{Bondi:1962px,Sachs:1962wk,Flanagan:2015pxa},
\begin{equation}
  \label{eq-bs-o}
  \begin{split}
    \ud s^2=&-\left(1-\frac{2\mathcal M}{r}\right)\ud u^2-2\ud u\ud r\\
    &+2\left\{\frac{\sD_Bc_A^B}{2}+\frac{1}{r}\left[\frac{2\mathcal N_A}{3}-\frac{\sD_A(c_C^Bc_B^C)}{16}\right.\right.\\
      &\left.\left.-\frac{c^A_B\sD^Cc^B_C}{2}\right]\right\}\ud u\ud\theta^A+r^2\left(\gamma_{AB}+\frac{c_{AB}}{r}\right)\ud\theta^A\ud\theta^B\\
    &+\cdots.
  \end{split}
\end{equation}
In this expression, $\mathcal M$ is the Bondi mass aspect, $c_{AB}$ is called the shear tensor, and $\mathcal N_A$ the angular momentum aspect.
With this metric ansatz, the vacuum Einstein's equation can be reduced to evolution equations of $\mathcal M$ and $\mathcal N_A$ in terms of $c_{AB}$, $\mathcal M$, and their various derivatives \cite{Barnich:2010eb,Flanagan:2015pxa,Madler:2016xju}.
Once the time dependence of $c_{AB}$ is specified, $\mathcal M$ and $\mathcal N_A$ are known, given their initial values, and the spacetime is determined.
A choice of tensors $\mathcal M$, $c_{AB}$ and $\mathcal N_A$ defines a Bondi frame \cite{Barnich:2010eb}.

The shear tensor $c_{AB}$ is intimately related to the radiative degrees of freedom in GR, as shown in some formal discussions \cite{Ashtekar:1981hw,Ashtekar:2018lor}.
This can also be easily seen by computing the geodesic deviation equation \cite{Wald:1984rg} for two adjacent test particles at a far distance from the source of gravity.
The 4-velocities of these test particles can be approximately given by $\pd_u+\cdots$, and one can orientate the test particles such that the deviation vector is in the radial direction, $V^A$.
So the geodesic deviation equation is
\begin{equation}
  \label{eq-gdv}
  \frac{\ud^2V^A}{\ud u^2}=2r\ddot c^{AB}V_B+\cdots.
\end{equation}
It is well-known that the interferometers detect the gravitational wave based on the geodesic deviation equation at the null infinity \cite{mtw}.
So the nonvanishing of $\ddot c_{AB}$ indicates the presence of the gravitational wave, and $c_{AB}$ contains the radiative degrees of freedom in general relativity.
In fact, if one defines the news tensor $N_{AB}\equiv\dot c_{AB}$, one usually calls a spacetime with $N_{AB}=0$ nonradiative or stationary.
For a stationary spacetime without any matter fields, it is possible to find a Bondi frame, such that \cite{Flanagan:2015pxa},
\begin{equation}
  \label{eq-vac-c}
  c_{AB}=\sD_{\langle A}\sD_{B\rangle}\Upsilon(\bm\theta),
\end{equation}
for some function $\Upsilon(\bm\theta)$ that is a linear combination of $Y^{\ell m}$ with $\ell\ge2$.
A choice of this function describes a vacuum gravitational state \cite{Ashtekar:1981hw,Strominger:2014pwa}.
Among the different Bondi frames, there is a canonical one, in which
\begin{subequations}
  \begin{gather}
    \mathcal M=\text{const.},\\
    c_{AB}=0,\\
    \mathcal N_A=\epsilon_{AB}\sD^B\Theta(\bm\theta),
  \end{gather}
\end{subequations}
with $(\sD^2+2)\Theta=0$ \cite{Flanagan:2015pxa}.

Although the spacetime described by Eq.~\eqref{eq-bs-o} generally does not possesses any global Killing vector field, asymptotically along the $r$ coordinate axis, the metric approaches the Minkowski, and it has certain approximate symmetries, i.e., the BMS symmetries \cite{Sachs1962asgr}.
These are diffeomorphisms or coordinate transformations that preserve the characteristics of the metric~\eqref{eq-bs-o}, i.e., \cite{Barnich:2010eb}
\begin{equation}
  \label{eq-bsc}
  g_{rr}=g_{rA}=0,\quad \det(r^{-2}g_{AB})=\det(\gamma_{AB}),
\end{equation}
and also the falloff behaviors explicitly displayed by Eq.~\eqref{eq-bs-o}.
An infinitesimal BMS transformation ($x^\mu\rightarrow x^\mu+\xi^\mu$) can be easily determined, given by \cite{Flanagan:2015pxa}
\begin{equation}
  \label{eq-bms-xi}
  \begin{split}
    \xi=&f\pd_u+\left(\mathcal Y^A-\frac{\sD^Af}{r}+\frac{c^{AB}\sD_Bf}{2r^2}\right)\pd_A\\
    &-\left(\frac{r}{2}\psi-\frac{\sD^2f}{2}\right)\pd_r+\cdots,
  \end{split}
\end{equation}
where $f=\alpha+u\psi/2$, $\alpha=\alpha(\bm\theta)$, $\psi=\sD_A\mathcal Y^A$, and $\mathcal Y^A=\mathcal Y^A(\bm\theta)$.
$\alpha$ parameterizes the so-called supertranslation, and $\mathcal Y^A$ generates the Lorentz transformation.
Let $n_i=(\sin\theta\cos\phi,\sin\theta\sin\phi,\cos\theta)$ with $i=1,2,3$ and $t^\mu=(t^0,t^i)$ represent a constant 4-vector field.
If $\alpha=t^0-t^in_i$, it generates the ordinary time ($t^0$) and space ($t^in_i$) translations.
A more general $\alpha$ is given by \cite{Flanagan:2015pxa}
\begin{equation}
  \label{eq-prop-al}
  \alpha=t^0-t^in_i+\sum_{\ell\ge2}\sum_{m=-\ell}^\ell\alpha_{\ell m}Y^{\ell m}(\bm\theta),
\end{equation}
with $Y^{\ell m}$ the spherical harmonics, and the last part represents a proper supertranslation.
Note that the first two terms are actually the linear combinations of $Y^{\ell m}$ with $\ell=0,1$.
$\mathcal Y^A$ can be written as
\begin{equation}
  \label{eq-y-dec}
  \mathcal Y^A=\omega^{0i}\sD^An_i-\omega^{ij}n_{[i}\sD^An_{j]},
\end{equation}
with $\omega^{\mu\nu}$ an antisymmetric constant tensor.
The first term generates the Lorentz boost, and the second the spatial rotation.
As discussed in the Introduction, super-Lorentz transformations are allowed in the extended or generalized BMS groups.
Their generators are given by more general expressions.
For example, in the generalized BMS group, $\mathcal Y^A=\sD^AY^{\ell m}+\epsilon^{AB}\sD_BY^{\ell m}$.

Under an infinitesimal asymptotic symmetry transformation \eqref{eq-bms-xi}, $\mathcal M$, $c_{AB}$ and $\mathcal N_A$ change according to \cite{Hou:2021oxe}
\begin{subequations}
  \label{eq-inf-bms-r}
  \begin{gather}
    \begin{split}
      \fL_\xi\mathcal M=&f\dot{\mathcal M}+\frac{1}{4}N^{AB}\sD_A\sD_Bf+\frac{1}{2}\sD_Af\sD_BN^{AB}\\
      &+\frac{3\psi}{2}\mathcal M+\mathcal Y^A\sD_A\mathcal M+\frac{1}{8}c^{AB}\sD_A\sD_B\psi,
    \end{split}\\
    \fL_\xi c_{AB}=fN_{AB}-2\sD_{\langle A}\sD_{B\rangle}f
    -\frac{\psi}{2}c_{AB}+\lie_{\mathcal Y}c_{AB},\label{eq-btf-c}\\
    \begin{split}
      \fL_\xi\mathcal N_A&=f\dot{\mathcal N}_A+\lie_{\mathcal Y}\mathcal N_A+\psi\mathcal N_A+3\mathcal M\sD_Af\\
      &+\frac{3}{4}(\sD_A\sD_Cc^C_B-\sD_B\sD_Cc^C_A+c_{AC}N^C_B)\sD^Bf,
    \end{split}
  \end{gather}
\end{subequations}
where $\lie_{\mathcal Y}$ denotes the Lie derivative on the 2-sphere.
Therefore, under an infinitesimal \textit{proper} supertranslation, a vacuum state transforms to a different one,
\begin{equation}
  c_{AB}\rightarrow c_{AB}-2\sD_{\langle A}\sD_{B\rangle}\alpha,
\end{equation}
that is, a time-independent $\chi(\bm\theta)$ changes to $\chi(\bm\theta)-2\alpha(\bm\theta)$, still time-independent.
Therefore, a proper supertranslation induces the vacuum transition, while a translation ($\alpha=t^0-t^in_i$) preserves it.

The Noether's theorem implies the existence of conserved charges.
The asymptotic charges associated with the generator $\xi^\mu$ can be derived based on certain covariant phase method \cite{Ashtekar:1981bq,Wald:1999wa,Barnich:2011mi,Flanagan:2015pxa,Alessio:2019cae}, and they are
\begin{subequations}
  \label{eq-bms-charges}
  \begin{equation}
    \label{eq-st-c}
    \mathcal Q_\alpha=\frac{1}{4\pi}\int\ud^2\bm\theta\sqrt\gamma\alpha \mathcal M,
  \end{equation}
  for the supertranslation $\alpha$, and
  \begin{equation}
    \label{eq-lo-c}
    \begin{split}
      \mathcal Q_{\mathcal Y}=&\frac{1}{8\pi}\int\ud^2\bm\theta\sqrt\gamma\mathcal Y^A\bigg[\mathcal N_A-u\sD_A\mathcal M\\
        &-\frac{1}{16}\sD_A(c_B^Cc^B_C)-\frac{1}{4}c_A^B\sD_Cc_B^C\bigg],
    \end{split}
  \end{equation}
\end{subequations}
for the super-Lorentz transformation $\mathcal Y^A$.
More specifically, if $\alpha=t^0=1$ \footnote{$t^0$ is said to be normalized.}, $\mathcal Q_{\alpha=1}$ is the Bondi mass, i.e., the total gravitational energy.
If $\alpha=n_i$, $\mathcal Q_{\alpha=n_i}$ gives the $i$-component of the spatial momentum.
For a proper supertranslation $\alpha=Y^{\ell m}$ with $\ell\ge2$, one gets the proper supertranslation charge $\mathcal Q_{\alpha=Y^{\ell m}}$, which may be named the supermomentum \cite{Ashtekar:1981bq,Flanagan:2015pxa,McCarthy1975bms,Barnich:2015uva}.
Similarly, one can use Eq.~\eqref{eq-lo-c} to compute the angular momentum  if $\mathcal Y^A=n_{[i}\sD^An_{j]}$, and the boost charge (i.e., the center-of-mass) if $\mathcal Y^A=\sD^An_i$.
For other forms of $\mathcal Y^A$, one can also obtained the super-Lorentz charges.
If there is the gravitational wave, these charges evolve with $u$ \cite{Hou:2021bxz},
\begin{subequations}
  \begin{gather*}
    \begin{split}
      & Q_\alpha(u_2)-Q_\alpha(u_1)\\
      =&\int_{u_1}^{u_2}\ud u\int\ud^2\theta\sqrt\gamma\alpha\left(\sD_A\sD_BN^{AB}+\frac{1}{2}N_A^BN^A_B\right),
    \end{split}\\
    \begin{split}
      &Q_{\mathcal Y}(u_2)-Q_{\mathcal Y}(u_1)\\
      =&\int_{u_1}^{u_2}\ud u\int\ud^2\theta\sqrt\gamma\bigg\{\frac{u\psi}{2}\left(\sD_A\sD_BN^{AB}+\frac{1}{2}N_A^BN^A_B\right)\\
      &+\frac{\mathcal Y^A}{4}\Big[N_B^C\sD_Ac_B^C
      -2\sD^B(N_B^Cc_{AC})-\langle c\leftrightarrow N\rangle\Big]\bigg\},
    \end{split}
  \end{gather*}
\end{subequations}
where $\langle c\rightarrow N\rangle$ means to exchange $c$ and $N$ in the previous terms in the squared brackets.
From these expressions, it is apparent that in a stationary spacetime with $N_{AB}=0$, $\mathcal Q_\alpha$ and $\mathcal Q_{\mathcal Y}$ are truly conserved.
They can be used to label different stationary spacetimes.
The changes in them indicate that the spacetime is physically distinct from the original one.

\subsection{The finite supertranslation}
\label{sec-f-st}

In this work, the finite supertranslation is needed.
To determine it, one shall make the following coordinate transformation,
\begin{subequations}
  \begin{gather}
    \hat u=u+\beta+\frac{\beta_1}{r}+\cdots,\\
    \hat r=r+\rho_0+\frac{\rho_1}{r}+\cdots,\\
    \hat\theta^A=\theta^A+\frac{\chi_1^A}{r}+\cdots,
  \end{gather}
\end{subequations}
with $\beta,\beta_1,\rho_0,\rho_1,\chi_1^A$ all functions of $(u,\bm\theta)$.
The form of this coordinate transformation is chosen such that in the hatted coordinates, the metric components $\hat g_{\mu\nu}$ satisfy the corresponding version of Eq.~\eqref{eq-bsc} and have the similar falloff behaviors to Eq.~\eqref{eq-bs-o}.
After some complicated mathematical manipulation, one obtains that \cite{Flanagan:2023jio}
\begin{subequations}
  \begin{gather}
    \dot\beta=0,\quad \beta_1=-\frac{1}{2}\sD_A\beta\sD^A\beta,\\
    \rho_0=\frac{1}{2}\sD^2\beta,\\
    \chi_1^A=-\sD^A\beta,
  \end{gather}
\end{subequations}
and $\rho_1$ is too complicated, which can be found in Ref.~\cite{Flanagan:2023jio}.
So a finite supertranslation is parameterized by a finite $\beta(\bm\theta)$, which can be written as a linear combination of spherical harmonics $Y^{\ell m}$ with $\ell\ge2$.
When $\beta$ is infinitesimal (like $\alpha$), this coordinate transformation gives rise to to Eq.~\eqref{eq-bms-xi} with $\mathcal Y^A=0$.
The transformation rules for $\mathcal M$, $c_{AB}$ and $\mathcal N_A$ are \cite{Flanagan:2023jio},
\begin{subequations}
  \begin{gather}
    \hat c_{AB}=c_{AB}-2\sD_{\langle A}\sD_{B\rangle}\beta,\label{eq-c-st}\\
    \begin{split}
      \hat{\mathcal M}=&\mathcal M+\frac{1}{2}\sD_AN^{AB}\sD_B\beta
      +\frac{1}{4}N^{AB}\sD_A\sD_B\beta\\
      &+\frac{1}{4}\dot N^{AB}\sD_A\beta\sD_B\beta,
    \end{split}\\
    \begin{split}
      \hat{\mathcal N}_A=&\mathcal N_A+3\mathcal M\sD_A\beta
      +\frac{3}{4}\sD_B\beta\big(\sD_A\sD^Cc^B_C-\sD^B\sD_Cc_A^C\\
      &+c_A^CN^B_C\big)+\frac{1}{2}\sD_A\beta\sD^B\beta\big(3\sD^CN_B^C+\dot N_{B}^C\sD_C\beta\big)\\
      &-\frac{1}{4}\sD_B\beta\sD^B\beta\big(3\sD_CN_A^C+\dot N^C_A\sD_C\beta\big).
    \end{split}
  \end{gather}
\end{subequations}
In addition, $\hat\gamma_{AB}=\gamma_{AB}$.
On the right-hand sides, $\gamma_{AB}$, $c_{AB}$, $\mathcal M$ and $\mathcal N_A$
are evaluated at $(u+\beta,\bm\theta)$, while on the left-hand side, their hatted counterparts are at $(u,\bm\theta)$.
One can check that these expressions are consistent with Eq.~\eqref{eq-inf-bms-r}.

According to Eq.~\eqref{eq-bms-charges}, the charges depend on the spacetime metric, characterized by $\mathcal M$, $c_{AB}$ and $\mathcal N_A$.
For a stationary spacetime, their transformations are simplified \cite{Flanagan:2023jio},
\begin{subequations}
  \begin{gather}
    \hat c_{AB}=c_{AB}-2\sD_{\langle A}\sD_{B\rangle}\beta,\\
    \hat{\mathcal M}=\mathcal M,\\
    \hat{\mathcal N}_A=\mathcal N_A+3\mathcal M\sD_A\beta
    +\frac{3}{2}\sD^B\beta\big(\sD_{[A}\sD^Cc_{B]C}\big).
  \end{gather}
\end{subequations}
Therefore, the conserved charges differ from the original ones according to
\begin{subequations}
  \begin{gather}
    \hat{\mathcal Q}_\alpha=\mathcal Q_\alpha,\\
    \begin{split}
      \hat{\mathcal Q}_{\mathcal Y}=&\mathcal Q_{\mathcal Y}
      +\frac{1}{8\pi}\int\ud^2\theta\sqrt\gamma \mathcal Y^A\bigg\{3\mathcal M\sD_A\beta\\
      &+\frac{3}{2}\sD^B\beta\sD_{[A}\sD^Cc_{B]C}+\frac{1}{2}\sD_{\langle A}\sD_{B\rangle}\beta\sD_Cc^{BC}\\
      &+\frac{1}{4}\sD_A[(c^{BC}-\sD^{\langle B}\sD^{C\rangle}\beta)\sD_{B}\sD_{C}\beta]\\
      &+\frac{1}{4}(c_{AB}-2\sD_{\langle A}\sD_{B\rangle}\beta)\sD^B(\sD^2+2)\beta
      \bigg\}.
    \end{split}
  \end{gather}
\end{subequations}
The supertranslation charge remains the same, because the supertranslation group is commutative, but the super-Lorentz charge changes.
Since these charges $\mathcal Q_\alpha$ and $\mathcal Q_{\mathcal Y}$ label physically distinguishable spacetimes, the supertranslation alters the physical state of the spacetime.

\section{Schwarzschild spacetime}
\label{sec-sch}

The Schwarzschild metric is one of the earliest nontrivial solutions to Einstein's equation \cite{Frolov:2011inbh}.
Its standard form possesses the time translation symmetry and the spatial rotation symmetry, explicitly.
So it describes the geometry outside of a spherically symmetry celestial object.
When the radius of this celestial object is small enough, it is a black hole, a highly compact object with the horizon, which is the boundary of no return.
It is generally believed to be the final state of a collapsing star.

\subsection{The standard Schwarzschild black hole}
\label{sec-sd-sch}

The metric of the standard Schwarzschild black hole is given by
\begin{equation}
  \label{eq-met-sch}
  \ud \mathring s^2=\bar g_{ab}\ud x^a\ud x^b+r^2\gamma_{AB}\ud\theta^A\ud\theta^B.
\end{equation}
Here, the first part of the metric can be written in three different coordinate systems in the following way,
\begin{equation}
  \label{eq-gab-c}
  \begin{split}
    \bar g_{ab}\ud x^a\ud x^b&=-F(r)\ud t^2+\frac{\ud r^2}{F(r)}\\
    &=-F(r)\ud u^2-2\ud u\ud r\\
    &=-F(r)\ud v^2+2\ud v\ud r,
  \end{split}
\end{equation}
where $F(r)=1-2M/r$, $u=t-r_*$, and $v=t+r_*$ with $r_*=r+2M\ln(r/2M-1)$.
$u$ is the retarded time, and $v$ is the advanced time.
These coordinates are suitable for describing the outgoing and incoming null radiations, respectively.
The second part of the metric \eqref{eq-met-sch} is conformal to the metric of the unit 2-sphere with the conformal factor $r^2$.
This particular form is taken from Ref.~\cite{Martel:2005ir}, which has been adopted to the gauge-invariant formalism reviewed in the next section.
The gauge-invariant formalism is very useful for the current work.

This form of the metric explicitly exhibits its symmetry.
There are four basic Killing vector fields,
\begin{subequations}
  \label{eq-sch-Killing}
  \begin{gather}
    \chi_t=\pd_t=\pd_u=\pd_v,\label{eq-sch-kt}\\
    \chi_x=-\sin\phi\pd_\theta-\cos\phi\cot\theta\pd_\phi,\label{eq-sch-kx}\\
    \chi_y=\cos\phi\pd_\theta-\sin\phi\cot\theta\pd_\phi,\label{eq-sch-ky}\\
    \chi_z=\pd_\phi.\label{eq-sch-kz}
  \end{gather}
\end{subequations}
The first vector $\chi_t$ generates the time translation symmetry, and it is null on the horizon.
The remaining vectors generate the spatial SO(3) rotations.
Their labels make good sense, as in the isotropic coordinates, the metric is \cite{Aichelburg:1970dh}
\begin{equation*}
  \label{eq-iso-sch}
  \ud \mathring s^2=-\left(\frac{1-M/2r}{1+M/2r}\right)^2\ud t^2+\left(1+\frac{M}{2r}\right)^4(\ud x^2+\ud y^2+\ud z^2),
\end{equation*}
with $r=\sqrt{x^2+y^2+z^2}$,
and once $x,y,z$ are expressed in terms of $r,\theta,\phi$ in the usual manner, the 3-metric in the last round brackets becomes $\ud r^2+r^2\gamma_{AB}\ud\theta^A\ud\theta^B$.
In addition to these continuous symmetries, there also exists the parity symmetry,
\begin{equation}
  \label{eq-def-p}
  \mathscr P: (u,r,\theta,\phi)\mapsto(u,r,\pi-\theta,\pi+\phi),
\end{equation}
and the metric remains the same.

The coordinate systems in Eq.~\eqref{eq-met-sch} are adopted to the symmetries of the metric.
This results in some important consequences.
In particular, for perturbations about the standard Schwarzschild metric, their equations can be simplified a lot, and the solutions can be derived with the method of the separation of variables.
For example, for a massless scalar perturbation $S$, the Klein-Gordon equation is
\begin{equation}
  \label{eq-sd-dal}
  \left[-2\pd_u\pd_r-\frac{2}{r}\pd_u+\frac{1}{r^2}\pd_rr^2F(r)\pd_r+\frac{1}{r^2}\sD^2\right]S=0.
\end{equation}
This equation is obviously invariant under the time translation, the spatial rotations and the parity transformation, which means that the d'Alembertian operator $\Box$ commutes with $\chi_t$, $\sD^2$, $\chi_i$ ($i=x,y,z$), and $\mathscr P$.
Here, $\chi_t$ and $\chi_i$ are used as the linear operators on a scalar field, which can be viewed as Lie derivatives.
Therefore, a solution $S$ can be separated into a radial part $\mathcal R(r)$, and a remaining part,  $e^{-i\omega t}Y^{\ell m}$, a common eigenfunction of operators $\chi_t$, $\sD^2$, $\chi_z$ and $\mathscr P$.
Its asymptotic behavior near the null infinity is
\begin{equation}
  \label{eq-scalar-null}
  S(u,r,\bm\theta)=\frac{1}{r}e^{-i\omega u}Y^{\ell m}+\cdots.
\end{equation}
Of course, a general solution is a linear combination of the above expression.
The equations for the vector and tensor perturbations are more complicated, but the similar simplifications also exist.
The metric perturbation will be considered more carefully and reviewed in Sec.~\ref{sec-bhper}.
The treatment of the vector perturbation can be found in Ref.~\cite{Berti:note}, for example.

Before discussing the supertranslated version of Eq.~\eqref{eq-met-sch}, let us inspect this metric more carefully.
By comparing Eq.~\eqref{eq-met-sch}  in $(u,r,\bm\theta)$ with Eq.~\eqref{eq-bs-o}, one can easily find out that the metric is actually in the Bondi gauge \cite{Bondi:1962px,Sachs:1962wk}.
So the black hole mass $M$ is the Bondi mass aspect, and the angular momentum aspect is zero.
The shear tensor also vanishes.
Thus this metric describes a gravitational vacuum state, and in particular, it is in the canonical Bondi frame.
In addition, the four Killing vector fields are also the elements of the BMS algebra.
$\chi_t$ is one of the supertranslation generators, corresponds to $\alpha=1$.
$\chi_i$'s are three of the Lorentz generators, corresponding to the second part of Eq.~\eqref{eq-y-dec} with the suitable choices of $\omega^{ij}$.
The Noether charges of these generators can be determined.
According to the previous  section, the only nontrivial charge is the total mass, $\mathring{\mathcal Q}_{\alpha=1}=M$, conjugate to $\chi_t$.

\subsection{Supertranslated Schwarzschild black holes}
\label{sec-st-sbh}

Ever since the program on the infrared structure of gravity renewed by Refs.~\cite{Strominger2014bms,Strominger:2014pwa,He:2014laa}, one starts to realize that a generic gravitational collapse may result in some stationary black hole spacetime that is supertranslated from the Kerr or Schwarzschild black hole.
In fact, as long as the energy flux of the collapsing matter contain components that are proportional to $Y^{\ell m}(\bm\theta)$ with $\ell\ge2$, the final stationary object would be in a nontrivial vacuum state with the soft hair \cite{Compere:2016hzt,Hawking:2016msc,Hawking:2016sgy,Chu:2018tzu,Strominger:2018inf}.
So in this subsection, let us consider such kind of final stationary states.

To get a supertranslated metric, one performs a finite supertranslation $\beta(\bm\theta)$, i.e.,
\begin{subequations}
  \label{eq-st-sbh}
  \begin{gather}
    \hat u=u+\beta-\frac{\sD_A\beta\sD^A\beta}{2r}+\cdots,\\
    \hat r=r+\frac{\sD^2\beta}{2}+\cdots,\\
    \hat \theta^A=\theta^A-\frac{\sD^A\beta}{r}+\cdots,
  \end{gather}
\end{subequations}
where it is natural to assume that the higher order terms in $1/r$ are independent of $u$.
Under this coordinate transformation, the metric, Eq.~\eqref{eq-met-sch}, becomes,
\begin{equation}
  \label{eq-st-sbh-0}
  \begin{split}
    \ud \hat{\mathring s}^2=&-\left(1-\frac{2M}{r}\right)\ud u^2-2\ud u\ud r\\
    &+2\bigg\{-\frac{\sD_A(\sD^2+2)\beta}{2}+\frac{1}{r}\bigg[2M\sD_A\beta\\
      &-\frac{(\sD_A\sD_{\langle B}\sD_{C\rangle}\beta)\sD^{\langle B}\sD^{C\rangle}\beta}{2}\bigg]\bigg\}\ud u\ud\theta^A\\
    &+r^2\left(\gamma_{AB}-\frac{2\sD_{\langle A}\sD_{B\rangle}\beta}{r}\right)\ud \theta^A\ud\theta^B+\cdots.
  \end{split}
\end{equation}
This metric is obviously different from the standard one \eqref{eq-met-sch}.
It is still explicitly $u$-independent, but it no longer explicitly possesses the rotational symmetry.
It actually has the  rotational symmetry implicitly, as the Killing equation is a tensor equation.
The four basic Killing vectors are,
\begin{subequations}
  \label{eq-kv-st-bh}
  \begin{gather}
    \hat \chi_t=\pd_{u},\\
    \begin{split}
      \hat \chi_x=&(\sin\phi\pd_\theta\beta+\cos\phi\cot\theta\pd_\phi\beta)\pd_{ u}\\
      &+\frac{1}{2}(\sin\phi\pd_\theta\sD^2\beta+\cos\phi\cot\theta\pd_\phi\sD^2\beta)\pd_{ r}\\
      &-\sin\phi\pd_{\theta}-\cos\phi\cot\theta\pd_{\phi}+\cdots,
    \end{split}\\
    \begin{split}
      \hat \chi_y=&-(\cos\phi\pd_\theta\beta-\sin\phi\cot\theta\pd_\phi\beta)\pd_{ u}\\
      &-\frac{1}{2}(\cos\phi\pd_\theta\sD^2\beta-\sin\phi\cot\theta\pd_\phi\sD^2\beta)\pd_{ r}\\
      &+\cos\phi\pd_{\theta}-\sin\phi\cot\theta\pd_{\phi}+\cdots,
    \end{split}\\
    \hat \chi_z=-\pd_\phi\beta\pd_{u}-\frac{1}{2}\pd_\phi\sD^2\beta\pd_r+\pd_{\phi}+\cdots.
  \end{gather}
\end{subequations}
Here, $\pd_u$ is still the exact symmetry generator, and for the three rotational Killing vectors $\hat\chi_i$, only leading order terms in $1/r$ are included.
So the timelike Killing vector $\hat\chi_t$ agrees with Eq.~\eqref{eq-sch-kt} for the standard Schwarzschild spacetime.
This is expected, as the supertranslated metric \eqref{eq-st-sbh-0} is time independent.
However, the Killing vectors for the spatial rotation are rather different.
One can find out that for the supertranslated black hole, $\hat\chi_ i$'s  contain the components in the $\pd_{u}$- and $\pd_r$-directions.
One can rewrite them in the following way,
\begin{equation}
  \label{eq-c-so3}
  \hat\chi_i=-(\chi_i^A\sD_A\beta)\pd_u-\frac{\chi_i^A\sD_A\sD^2\beta}{2}\pd_r+\chi_i+\cdots.
\end{equation}
Since $\sD^2(\chi_i^A\sD_A\beta)=\chi_i^A\sD_A\sD^2\beta$ with $\sD^{(A}\chi_i^{B)}=0$ and $\sD^2\chi_i^A=-\chi_i^A$, it is clear that they now take the form of a generic BMS generator, referring to Eq.~\eqref{eq-bms-xi} and Ref.~\cite{Geroch1977}.
One can remove the $u$- and $r$-components of $\hat\chi_i$'s by adding to them the supertranslation generator  with $\alpha=\chi_i^A\sD_A\beta$, and $\hat\chi_i$'s become
\begin{equation}
  \hat\chi_i\rightarrow\hat\chi'_i= \chi_i+\cdots.
\end{equation}
Since the added supertranslation generator is not linear in $\hat\chi_t$, $\hat\chi'_i$'s cease to be Killing vectors of the supertranslated metric, so do $\chi_i$'s.

The supertranslated metric \eqref{eq-st-sbh-0} is not explicitly parity invariant, either.
Like the rotational symmetry, a hidden ``parity transformation'' $\hat{\mathscr P}$ may exist such that Eq.~\eqref{eq-st-sbh-0} is invariant.
It is difficult to determine $\hat{\mathscr P}$, given the disappearance of the explicit rotational symmetry of Eq.~\eqref{eq-st-sbh-0}.
But one may define an asymptotic $\hsP$ near the null infinity ($r\rightarrow+\infty$).
For that, one shall go back to the standard case and gain some ideas by digesting how to associate the rotation symmetry with the parity $\sP$.
To get $\sP$ as defined in Eq.~\eqref{eq-def-p} for the standard metric, one can move any event $(u_0,r_0,\theta_0,\phi_0)$ along the integral curve $\Gamma(\zeta)$ of the Killing vector field $\chi_0=\sin\phi_0\chi_x+\cos\phi_0\chi_y$ with $\Gamma(0)=(u_0,r_0,\theta_0,\phi_0)$.
One can check that $\chi_0$ generates the rotation around a direction perpendicular to the surface containing the $z$-axis and the radial direction determined by $(\theta_0,\phi_0)$.
$\Gamma(\zeta)$ lies on the 2-sphere of radius $r_0$ at any time $u_0$.
In fact, it is a great circle.
When $\zeta=\pi$, $(u_0,r_0,\theta_0,\phi_0)$ is moved to its ancipital point $\Gamma(\pi)=(u_0,r_0,\pi-\theta_0,\pi+\phi_0)$, that is, $\sP(u_0,r_0,\theta_0,\phi_0)=\Gamma(\pi)$.
Similarly, to define an asymptotic parity $\hsP$ for the supertranslated metric, one still starts with an arbitrary event, say $(u_0,r_0,\theta_0,\phi_0)$ with a very large $r_0$, and moves it along the integral curve $\hat\Gamma(\zeta)$ of $\hat\chi_0=\sin\phi_0\hat\chi_x+\cos\phi_0\hat\chi_y$.
It shall be emphasized that the exact Killing vectors $\hat\chi_i$'s for the supertranslated metric shall be used here.
Then, one defines $\hsP$ to be the mapping $\hsP(u_0,r_0,\theta_0,\phi_0)\equiv\hat\Gamma(\pi)$.
After some calculation, one can show that
\begin{equation}
  \label{eq-def-p-st}
  \hsP:\left[
    \begin{array}{c}
      u      \\
      r      \\
      \theta \\
      \phi
    \end{array}
    \right]\mapsto \left[
    \begin{array}{c}
      u+\beta(\bm\theta)-\beta(\sP\bm\theta)                                             \\
      \displaystyle r+\frac{\sD^2\beta(\bm\theta)}{2}-\frac{\sD^2\beta(\sP\bm\theta)}{2} \\
      \pi-\theta                                                                         \\
      \pi+\phi
    \end{array}
    \right]+\cdots,
\end{equation}
where $\bm\theta=(\theta,\phi)$ and $\sP\bm\theta=(\pi-\theta,\pi+\phi)$ to make the expressions less cluttered.
One can show that under this transformation, the leading order piece of Eq.~\eqref{eq-st-sbh-0}, i.e., the Minkowski metric, is invariant.
We will call $\hsP$ the parity operator for the supertranslated Schwarzschild spacetime.
Although this operator is defined asymptotically ($r\rightarrow+\infty$), it is useful to organize the perturbations of the supertranslated black hole later.

Now, one can read off the Bondi mass aspect, the shear tensor and the angular momentum aspect of the supertranslated metric.
By Eq.~\eqref{eq-st-sbh-0},  the Bondi mass aspect is still $M$, but the shear tensor and the angular momentum aspect are
\begin{subequations}
  \begin{gather}
    \hat{\mathring c}_{AB}=-2\sD_{\langle A}\sD_{B\rangle}\beta,\label{eq-st-va-c}\\
    \hat{\mathring{\mathcal N}}_A=2M\sD_A\beta,
  \end{gather}
\end{subequations}
respectively.
The form of $\hat{\mathring c}_{AB}$ suggests that the metric~\eqref{eq-st-sbh-0} still describes a gravitational vacuum state.
However, the asymptotic charges are different from those of the standard Schwarzschild spacetime.
The nontrivial conserved charges include the total mass,
\begin{subequations}
  \begin{equation}
    \hat{\mathring{\mathcal Q}}_{\alpha=1}=M,
  \end{equation}
  and the super-Lorentz charges,
  \begin{equation}
    \begin{split}
      \hat{\mathring{\mathcal Q}}_{\mathcal Y}=\frac{1}{16\pi}&\int\ud^2\theta\sqrt\gamma\mathcal Y^A\big[6M\sD_A\beta\\
        &-(\sD_A\sD_{\langle B}\sD_{C\rangle}\beta)\sD^{\langle B}\sD^{C\rangle}\beta\\
        &-(\sD_{\langle A}\sD_{B\rangle}\beta)\sD^B(\sD^2+2)\beta\big].
    \end{split}
  \end{equation}
\end{subequations}
In particular, if $\mathcal Y^A=\sD^AY^{1m}$ \footnote{Note that here $\ell=1$.}, the above equation gives the center-of-mass;
if $\mathcal Y^A=\chi_i^A$, one obtains the angular momentum in the $i$-direction.
That is, although one gets Eq.~\eqref{eq-st-sbh-0} via supertranslating the standard Schwarzschild metric, it has nonvanishing center-of-mass, angular momentum, and other super-Lorentz charges.
This implies that the supertranslated Schwarzschild metric describes a physically different vacuum state from the standard Schwarzschild spacetime.
One thus expects that there might also be some differences in their perturbations to be elaborated later.

Now, let us compare our supertranslated metric \eqref{eq-st-sbh-0} with Eq.~\eqref{eq-st-sch-s}.
They look differently.
However, they can be shown to be equal to each other near the null infinity.
Indeed, one can first expand the components of Eq.~\eqref{eq-st-sch-s} in powers of $1/\rho$, and then, replace $t$ by $u+\rho+\cdots$.
The resulting expression is similar to Eq.~\eqref{eq-st-sbh-0}, as long as one identifies $\rho$ as $r$.
In Ref.~\cite{Compere:2016hzt}, many kinematic properties of Eq.~\eqref{eq-st-sch-s} were studied, including the positions of horizon and the so-called supertranslation horizon, and the properties of the supertranslation horizon.
Since in the current work, the supertranslated Schwarzschild metric is obtained only near the null infinity, these kinematics may not be investigated.

Obviously, the coordinate system of Eq.~\eqref{eq-st-sbh-0} is not compatible with the symmetries of the metric.
So one expects that the equations of motion of various fields are much more complicated.
Again, take the scalar field $S$ for example.
The supertranslated d'Alembertian operator is
\begin{equation}
  \begin{split}
    \hat\Box=&-2\pd_u\pd_r-\frac{2}{r}\pd_u+\frac{1}{r^2}\pd_rr^2F(r)\pd_r+\frac{1}{r^2}\sD^2\\
    &-\frac{1}{r^2}\bigg[\sD^A(\sD^2+2)\beta\pd_A\pd_r+\frac{1}{2}\sD^2(\sD^2+2)\beta\pd_r\\
      &-\frac{1}{4}\sD_{\langle A}\sD_{B\rangle}\beta\sD^A\sD^B\beta(\pd_r-2\pd_u)\pd_r\bigg]+\cdots,
  \end{split}
\end{equation}
where the first line agrees with Eq.~\eqref{eq-sd-dal} for the standard Schwarzschild metric.
The presence of $\beta(\bm\theta)$ in the subsequent lines implies that $\hat\Box$ no longer commutes with $\sD^2$ or $\chi_i$'s.
Since $\hat\Box$ commutes with $\hat\chi_t$, one can separate $S$ such that $S=e^{-i\omega u}\mathcal S(r,\bm\theta)$ for some function $\mathcal S(r,\bm\theta)$.
It may be impossible to further separate $\mathcal S$, if $\beta$ does not have a simple enough functional form.
Luckily, the invariance of the Klein-Gordon equation under a coordinate transformation suggests that the solution to the supertranslated Klein-Gordon equation can be obtained by supertranslating the standard solution,
\begin{equation}
  \hat S(u,r,\bm\theta)=S(u+\beta+\cdots,r+\sD^2\beta/2+\cdots,\bm\theta+\cdots),
\end{equation}
and its leading order part near the null infinity is rather simple, i.e., a linear combination of
\begin{equation}
  \label{eq-lo-sp}
  \frac{1}{r}\exp[-i\omega (u+\beta)]Y^{\ell m}.
\end{equation}
%
It is easy to check that it is not only the eigenfunction of $\hat\chi_t$, but also the eigenfunction of $\hat\sD^2$, $\hat\chi_z$, and $\hsP$, with eigenvalues $-\ell(\ell+1)$, $im$, and $(-1)^{\ell}$, respectively.
Here, the definitions of $\hat\sD^2$ and $\hat\chi_z$ are given in Appendix~\ref{sec-aop}.
So one can get the supertranslated scalar perturbation by shifting the retarded time $u\rightarrow u+\beta$ at the null infinity.
Once one fixes the leading order piece of $\hat S$ in $1/r$, one can determined the higher order terms of $\hat S$ using its equation of motion.
The treatment of the tensor perturbation is similar as presented in Sec.~\ref{sec-stper}.

\subsubsection{Supertranslated metric with cylindrical symmetry}
\label{sec-azi}

There is a family of special choices of $\beta(\bm\theta)$ that do not break the whole rotational symmetry.
One obvious choice is
\begin{equation}
  \label{eq-azi-z}
  \beta=\beta(\theta),
\end{equation}
so the rotational symmetry around the $z$-axis still holds \emph{explicitly} for the supertranslated Schwarzschild black hole, i.e.,
\begin{equation}
  \hat\chi_z=\pd_\phi,
\end{equation}
exactly same as $\chi_z$ for the standard case.
Another less obvious choice is
\begin{equation}
  \label{eq-azi-x}
  \beta=\beta(\sin\theta\cos\phi),
\end{equation}
which is invariant under the rotation around the $x$-axis, and in the supertranslated spacetime,
\begin{equation}
  \hat\chi_x=\chi_x,
\end{equation}
too.
Similarly, if $\beta=\beta(\sin\theta\sin\phi)$, the Killing vector $\hat\chi_y=\chi_y$ for the rotation around the $y$-axis remains the same.

A more general form of $\beta(\bm\theta)$ that is invariant under a rotation around the direction described by the angles $(\iota,o)$ is given by
\begin{equation}
  \label{eq-azi-0}
  \beta=\beta(\mathcal B),\quad \mathcal B=\cos\theta\cos\iota+\cos(o-\phi)\sin\theta\sin\iota.
\end{equation}
In fact, the Killing vector field that generates the rotation around $(\iota,o)$ is
\begin{equation}
  \begin{split}
    \chi_{a}=&\sin\iota\cos o\chi_x+\sin\iota\sin o\chi_y+\cos\iota\chi_z\\
    =&\sin\iota\sin(o-\phi)\pd_\theta\\
    &-[\cos(o-\phi)\cot\theta\sin\iota-\cos\iota]\pd_\phi,
  \end{split}
\end{equation}
in the standard Schwarzschild spacetime,
and it can be checked that,
\begin{equation}
  \lie_{\chi_a}\beta=\lie_{\chi_a}\mathcal B=0.
\end{equation}
Therefore, if the standard Schwarzschild black hole is supertranslated  by $\beta(\mathcal B)$, the resulting spacetime would have $\hat\chi_a=\chi_a$ as the generator for the SO(2) symmetry round the axis defined by $(\iota,o)$.

As a final remark, one shall clearly distinguish the supertranslated Schwarzschild metric with $\beta=\beta(\theta)$ from the Kerr metric, although both are symmetric around the $z$-axis.
After rewriting the Kerr metric in the Bondi-Sachs coordinates \cite{Fletcher2003bs,Bishop:2006kerr,Bai:2007rs,Wu:2008yi}, one can verify that the Kerr metric is not supertranslated.

\section{Black hole perturbation theory}
\label{sec-bhper}

The black hole perturbation theory has been developed for a long time, initiated by Refs.~\cite{Regge:1957td,Vishveshwara:1970cc,Zerilli:1970wzz}.
It takes advantage of the (exact) symmetries of the background spacetime, and the linearized Einstein's equation can be simplified greatly.
For the perturbations above the standard Schwarzschild black hole, the symmetries include the time translation ($\chi_t$) and the spatial rotations ($\chi_i$'s).
Thus, the metric perturbation can be Fourier transformed and decomposed using spherical harmonics.
The perturbed Einstein's equation can be rewritten as two sets of master equations governing the radial behaviors of the Regge-Wheeler and Zerilli functions (or their variants).
Certain gauge-invariant formalism has also been invented \cite{Gerlach:1979rw,Ripley:2017kqg,Clarkson:2002jz,Martel:2005ir}.
This formalism is very suitable for current work, so it will be briefly reviewed in this section.

Let $p_{\mu\nu}$ represent the perturbation to the background metric \eqref{eq-met-sch}.
Since the metric \eqref{eq-met-sch} enjoys the spherical symmetry, it is convenient to expand the components $p_{ab}, p_{aA}, p_{AB}$ in terms of the spherical harmonics $Y^{\ell m}(\theta)$, and their vectorial and tensorial versions.
The vectorial spherical harmonics include
\begin{equation}
  Y_A^{\ell m}=\sD_AY^{\ell m},\quad X_A^{\ell m}=-\epsilon_A{}^B\sD_BY^{\ell m},
\end{equation}
which are the even-parity and odd-parity harmonics, respectively.
The tensorial spherical harmonics can be constructed in the following way,
\begin{equation}
  \label{eq-tensph}
  Y_{AB}^{\ell m}=\sD_{\langle A}\sD_{B\rangle}Y^{\ell m},\quad
  X_{AB}^{\ell m}=-\epsilon_{(A}{}^C\sD_{B)}\sD_CY^{\ell m},
\end{equation}
which are traceless.
There are also tensorial harmonics, $\gamma_{AB}Y^{\ell m}$, which represent the traces of  tensors on the 2-sphere.
Under the parity transformation, $Y_{AB}^{\ell m}$ and $\gamma_{AB}Y^{\ell m}$ transform in the even-parity representation, while $X_{AB}^{\ell m}$ the odd-party.
Finally, $Y^{\ell m}$ are even.
These spherical harmonics are eigenfunctions of $\sD^2$ and $\chi_z$.
The eigenvalues are shown in Table~\ref{tab-ev-ef}.
\begin{table}[h]
  \centering
  \begin{tabular}{c|c|c|c}
    \hline\hline
    spherical harmonics & $Y^{\ell m}$              & $Y_A^{\ell m},X_A^{\ell m}$ & $Y_{AB}^{\ell m},X_{AB}^{\ell m}$ \\
    \hline
    $\sD^2$             & $-\ell(\ell+1)$           & $1-\ell(\ell+1)$            & $4-\ell(\ell+1)$                  \\
    \hline
    $\chi_z$            & \multicolumn{3}{c}{$-im$}                                                                   \\
    \hline
  \end{tabular}
  \caption{Eigenvalues of various spherical harmonics for operators $\sD^2$ and $\chi_z$.}
  \label{tab-ev-ef}
\end{table}

The perturbation $p_{\mu\nu}$ can thus be categorized into even and odd parity types, also known as polar and axial, respectively.
The even-parity perturbation is given by
\begin{subequations}
  \label{eq-e-p-he}
  \begin{gather}
    p_{ab}=\sum_{\ell m}h_{ab}^{\ell m}Y^{\ell m},\\
    p_{aA}=\sum_{\ell m}j_a^{\ell m}Y_A^{\ell m},\\
    p_{AB}=r^2\sum_{\ell m}(K^{\ell m}\gamma_{AB}Y^{\ell m}+G^{\ell m}Y^{\ell m}_{AB}),
  \end{gather}
\end{subequations}
where $h_{ab}^{\ell m},j_a^{\ell m},K^{\ell m}$, and $G^{\ell m}$ are functions of $(t,r)$ or $(u,r)$.
For the odd-parity perturbation, one has
\begin{subequations}
  \label{eq-o-p-he}
  \begin{gather}
    p_{ab}=0,\label{eq-o-p-he-1}\\
    p_{aA}=\sum_{\ell m}h_a^{\ell m}X_B^{\ell m},\label{eq-o-p-he-2}\\
    p_{AB}=\sum_{\ell m}h_2^{\ell m}X_{AB}^{\ell m}.\label{eq-o-p-he-3}
  \end{gather}
\end{subequations}
Similarly, $h_a^{\ell m}$ and $h_2^{\ell m}$ are another set of functions of $(t,r)$ or $(u,r)$.
In these expressions, $\ell\ge2$.

By studying the gauge transformation properties of the above expansion coefficients under an arbitrary infinitesimal coordinate transformation, one can construct the following gauge-invariant quantities \cite{Martel:2005ir},
\begin{subequations}
  \begin{gather}
    \tilde h_{ab}^{\ell m}=h^{\ell m}_{ab}-\bar\nabla_aE_b^{\ell m}-\bar\nabla_bE_a^{\ell m},\\
    \tilde K^{\ell m}=K^{\ell m}+\frac{\ell(\ell+1)}{2}G^{\ell m}-\frac{2}{r}r^aE_a^{\ell m},
  \end{gather}
  with  $E_a^{\ell m}=j_a^{\ell m}-r^2\bar\nabla_aG^{\ell m}/2$
  for the even-parity perturbation, and
  \begin{equation}
    \tilde h_a^{\ell m}=h^{\ell m}_a-\frac{1}{2}\bar\nabla_ah_2^{\ell m}+\frac{1}{2}r_ah_2^{\ell m},
  \end{equation}
\end{subequations}
for the odd-parity perturbation.
In these expressions, $r_a=\pd r/\pd x^a$, and $r^a=\bar g^{ab}r_b$.
Then, the perturbed Einstein's equation can be rewritten as
\begin{subequations}
  \begin{equation}
    \label{eq-ze}
    (-\pd_t^2+\pd_{r_*}^2-V_e)\Psi_e^{\ell m}=0,
  \end{equation}
  for the even-parity perturbation, and
  \begin{equation}
    \label{eq-rw}
    (-\pd_t^2+\pd_{r_*}^2-V_o)\Psi_o^{\ell m}=0,
  \end{equation}
\end{subequations}
for the odd-parity perturbation.
The original perturbed Einstein's equation is too complicated to be presented, which can be found in Ref.~\cite{Martel:2005ir}.
$\Psi_e^{\ell m}$ and $\Psi_o^{\ell m}$ are the Zerilli-Moncrief \cite{Lousto:1996sx} and Cunningham-Price-Moncrief \cite{Gerlach:1979rw,Cunningham:1978zfa} functions, respectively, given by,
\begin{gather*}
  \Psi_e^{\ell m}=\frac{2r}{\ell(\ell+1)}\left[\tilde K^{\ell m}+\frac{2}{\Lambda}r^a(r^b\tilde h_{ab}^{\ell m}-r\bar\nabla_a\tilde K^{\ell m})\right],\\
  \Psi_o^{\ell m}=\frac{2r}{(\ell-1)(\ell+2)}\bar\epsilon^{ab}\left(\bar\nabla_a\tilde h_b^{\ell m}-\frac{2}{r}r_a\tilde h_b^{\ell m}\right),
\end{gather*}
where $\Lambda=(\ell-1)(\ell+2)+6M/r$.
Their respective potentials are
\begin{gather*}
  V_e=\frac{F(r)}{\Lambda^2}\left[\mu^2\left(\frac{\mu+2}{r^2}+\frac{6M}{r^3}\right)+\frac{36M^2}{r^4}\left(\mu+\frac{2M}{r}\right)\right],\\
  V_o=F(r)\left[\frac{\ell(\ell+1)}{r^2}-\frac{6M}{r^3}\right],
\end{gather*}
with $\mu=(\ell-1)(\ell+2)$.
So these potentials are functions of $\ell$, although not explicitly labeled.
Neither of them depends on $m$, reflecting the spherical symmetry of the background metric.
Equation~\eqref{eq-ze} is called the Zerilli equation, and \eqref{eq-rw} is the Regge-Wheeler equation \cite{Martel:2005ir}.

Once suitable boundary conditions are specified, one can solve Eqs.~\eqref{eq-ze} and \eqref{eq-rw}, using Wentzel–Kramers–Brillouin method \cite{Schutz:1985km}, Leaver's method \cite{Leaver:1985ax}, or numerically.
Since both $V_e$ and $V_o$ decay to zero as $r\rightarrow+\infty$ and $r\rightarrow 2M$, one may impose the following boundary conditions,
\begin{equation}
  \label{eq-bdyc-sc}
  \Psi_{e/o}^{\ell m}=\left\{
  \begin{array}{cc}
    e^{-i\omega v}+\mathcal R^\ell_{e/o}e^{-i\omega u}, & r\rightarrow +\infty \\
    \mathcal T^\ell_{e/o}e^{-i\omega v},                & r\rightarrow 2M
  \end{array}
  \right.,
\end{equation}
for a scattering process, in which an incoming wave $e^{-i\omega v}$ is partially reflected by the potential barrier $V_{e/o}$ with the reflection coefficient $\mathcal R_{e/o}^\ell$, and partially transmitted with the transmission coefficient $\mathcal T^\ell_{e/o}$.
There also exist other boundary conditions in literature \cite{Futterman:1988ni}.
For the scattering problem, $\omega$ is real.
If one imposes
\begin{equation}
  \label{eq-bdyc-qnm}
  \Psi_{e/o}^{\ell m}=\left\{
  \begin{array}{cc}
    e^{i\omega u}, & r\rightarrow+\infty \\
    e^{i\omega v}, & r\rightarrow 2M
  \end{array}
  \right.,
\end{equation}
so that the energy carried by the wave escapes into the infinity, and enters the horizon,
one can obtained the quasinormal modes.
The energy outside of the horizon is decaying, so $\omega$ is complex.
Since the effective potentials $V_{e/o}$ depend on $\ell$, $\omega$ is a function of $\ell$.
In addition, for a specific $\ell$, $\omega$ may take multiple values, corresponding to  different overtone indices $n\ge0$.
So $\omega$ can be labeled as $\omega_{n\ell}$.
$\omega^2_{n\ell}$ can be viewed as the eigenvalue of the linear operators $\pd^2_{r_*}-V_{e/o}$ with the eigenfunctions $\Psi_{e/o}^{n\ell m}$, where the overtone index is added.
$\Psi_{e/o}^{n\ell m}$ are also the eigenfunctions of the Killing vector $\chi_t$ with the eigenvalue $i\omega_{n\ell}$.
As well known, the two wave equations~\eqref{eq-ze} and \eqref{eq-rw} share the same spectra.
This is the isospectrality that has been proved rigorously \cite{Chandrasekhar:1985kt}.
The quasinormal modes will be the further discussed below.

\section{The standard perturbations in Bondi gauge}
\label{sec-tbs}

Although the background metric \eqref{eq-met-sch} in $(u,r,\theta^A)$ is in the Bondi-Sachs gauge, the perturbed metric $\mathring g_{\mu\nu}+p_{\mu\nu}$ may not.
Thanks to the gauge-invariant formalism reviewed in Section~\ref{sec-bhper}, one can easily impose the Bondi-Sachs conditions, i.e.,
\begin{equation}
  \label{eq-bs-p}
  p_{rr}=p_{rA}=0,\quad \gamma^{AB}p_{AB}=0,
\end{equation}
without affecting the Zerilli or Regge-Wheeler equations.
In the following, the explicit form of $p_{\mu\nu}$ will be determined and expressed as a series expansion in inverse powers of $r$.
Since there are even- and odd-parity perturbations, one would like to compute their expressions separately, inspired by Ref.~\cite{Martel:2005ir}.

\subsection{Even-parity perturbations}
\label{sec-e-bs}

For the even-parity perturbation, the Bondi gauge conditions \eqref{eq-bs-p} imply that
\begin{equation}
  h_{rr}^{\ell m}=j_r^{\ell m}=K^{\ell m}=0,
\end{equation}
referring to Eq.~\eqref{eq-e-p-he}.
The remaining expansion coefficients in Eq.~\eqref{eq-e-p-he} can be further expanded in inverse powers of $r$,
\begin{subequations}
  \label{eq-e-p-rexp}
  \begin{gather}
    h_{uu}^{\ell m}=\sum_{k=1}\frac{\mathcal M^{\ell m}_k}{r^k},\quad h_{ur}^{\ell m}=\sum_{k=1}\frac{\mathcal B^{\ell m}_k}{r^k},\\
    j_u^{\ell m}=\sum_{k=0}\frac{\mathcal N^{\ell m}_k}{r^k},\\
    G^{\ell m}=\sum_{k=1}\frac{\mathcal C^{\ell m}_k}{r^k},
  \end{gather}
\end{subequations}
where $\mathcal M_k,\mathcal B_k,\mathcal N_k$, and $\mathcal C_k$ are functions of $u$.
Substituting Eq.~\eqref{eq-e-p-rexp} into the perturbed Einstein's equation, one finds out that
\begin{subequations}
  \begin{gather}
    \mathcal M_0^{\ell m}=\frac{(\ell+2)!}{4(\ell -2)!}\mathcal C_1^{\ell m},\quad \mathcal B^{\ell m}_1=0,\\
    \mathcal N^{\ell m}_0=-\frac{(\ell-1)(\ell+2)}{4}\mathcal C^{\ell m}_1,\\
    \pd_u\mathcal N^{\ell m}_1=\frac{(\ell+2)!}{12(\ell-2)!}\mathcal C^{\ell m}_1,\label{eq-evo-n1}
  \end{gather}
\end{subequations}
where only leading order expansion coefficients are presented, sufficient for the following discussion.
Once one knows the functional dependence of $\mathcal C^{\ell m}_1$ with the retarded $u$, one can integrate Eq.~\eqref{eq-evo-n1} to determine $\mathcal N^{\ell m}_1$.
$\mathcal C^{\ell m}_1$ is not constrained by the Einstein's equation, which represents the radiative degree of freedom of the theory.
Indeed, it can be shown that
\begin{equation}
  \Psi_e^{\ell m}=\mathcal C^{\ell m}_1+\order{\frac{1}{r}}.
\end{equation}
In addition, it contributes to the shear tensor,
\begin{equation}
  \label{eq-shear-e-ss-0}
  c_{AB}=\Psi_{e,0}^{\ell m}(u)Y_{AB}^{\ell m},
\end{equation}
where $\Psi_{e,0}^{\ell m}$ stands for the leading order part, i.e., $\mathcal C^{\ell m}_1$, and $\ell\ge2$.
Of course, for a general solution, one has to sum over $\ell\ge2$ and $m\in[-\ell,\ell]$.

$\mathcal M^{\ell m}_0$ is the correction to the Bondi mass aspect, and $\mathcal N^{\ell m}_1$ gives the angular momentum aspect, i.e.,
\begin{gather}
  \mathcal M=M+\frac{(\ell+2)!}{4(\ell-2)!}\Psi_{e,0}^{\ell m}Y^{\ell m},\\
  \mathcal N_A=\frac{3}{2}\mathcal N_1^{\ell m}Y_A^{\ell m}.
\end{gather}
One can check that these results are consistent with those in Ref.~\cite{Blanchet:2020ngx}.
Therefore, for a perturbed Schwarzschild black hole spacetime, its Bondi mass is no longer a constant, but oscillating.
It also has an oscillating angular momentum aspect.
According to Eq.~\eqref{eq-bms-charges}, the nontrivial charges include not only the total Bondi mass $\mathcal Q_{\alpha=1}=M$, but also the (proper) supertranslation charges
\begin{subequations}
  \begin{equation}
    \mathcal Q_{\alpha=Y^{\ell m}}=\frac{(-1)^m}{16\pi}\frac{(\ell+2)!}{(\ell-2)!}\Psi^{\ell,-m}_{e,0}.
  \end{equation}
  To compute the super-Lorentz charges, one can set $\mathcal Y^A=\sD^AY^{\ell m}$ to compute the super-boost charge, given by
  \begin{equation}
    \mathcal Q_{\sD^AY^{\ell m}}=(-1)^m\frac{1}{8\pi}\left[3\mathcal N_1^{\ell,-m}-u\frac{(\ell+2)!}{(\ell-2)!}\Psi^{\ell,-m}_{e,0}\right],
  \end{equation}
  and the super angular momenta vanish ($\mathcal Y^A=\epsilon^{AB}\sD_BY^{\ell m}$) at the linear order in the metric perturbation.
\end{subequations}

\subsection{Odd-parity perturbations}
\label{sec-o-bs}

For the odd-parity perturbation, Eqs.~\eqref{eq-o-p-he-1} and \eqref{eq-o-p-he-3} automatically satisfy the Bondi gauge conditions \eqref{eq-bs-p}.
So one has to require
\begin{equation}
  h_r^{\ell m}=0,
\end{equation}
and expand
\begin{subequations}
  \begin{gather}
    h_u^{\ell m}=\sum_{n=0}\frac{\mathcal N'^{\ell m}_k}{r^k},\\
    h_2^{\ell m}=\sum_{n=-1}\frac{\mathcal C'^{\ell m}_k}{r^k}.
  \end{gather}
\end{subequations}
After solving the perturbed Einstein's equation, one obtains  the following leading order solutions,
\begin{subequations}
  \begin{gather}
    \mathcal N'^{\ell m}_0=-\frac{(\ell-1)(\ell+2)}{4}\mathcal C'^{\ell m}_{-1},\\
    \pd_u\mathcal N'^{\ell m}_1=-\frac{(\ell+2)!}{12(\ell-2)!}\mathcal C'^{\ell m}_{-1}.\label{eq-evo-n1p}
  \end{gather}
\end{subequations}
Like $\mathcal C^{\ell m}_1$ in the previous subsection, $\mathcal C'^{\ell m}_{-1}$ is not constrained by the equations of motion.
By integrating Eq.~\eqref{eq-evo-n1p}, $\mathcal N'^{\ell m}_1$ can be obtained provided $\mathcal C'^{\ell m}_{-1}=\mathcal C'^{\ell m}_{-1}(u)$ is given.
It is easy to check that
\begin{equation}
  \Psi^{\ell m}_o=\mathcal C'^{\ell m}_{-1}+\order{\frac{1}{r}},
\end{equation}
and the shear tensor is
\begin{equation}
  \label{eq-shear-o-ss-0}
  c_{AB}=\Psi_{o,0}^{\ell m}(u)X_{AB}^{\ell m}.
\end{equation}
Again, here, $\Psi_{o,0}^{\ell m}$ is the leading order term, $\mathcal C'^{\ell m}_{-1}$, $\ell\ge2$, and no summation is performed.

Due to Eq.~\eqref{eq-o-p-he-1}, there is no correction to the Bondi mass aspect from the odd-parity perturbation, and the angular momentum aspect becomes
\begin{equation}
  \mathcal N_A=\frac{3}{2}\mathcal N'^{\ell m}_1X_A^{\ell m}.
\end{equation}
Therefore, the nontrivial asymptotic charges include the total Bondi mass $M$, and the super angular momenta ($\mathcal Y^A=\epsilon^{AB}\sD_BY^{\ell m}$), given by
\begin{equation}
  \mathcal Q_{\epsilon^{AB}\sD_BY^{\ell m}}=(-1)^{m+1}\frac{3}{8\pi}\mathcal N'^{\ell,-m}_1.
\end{equation}
The angular momenta are functions of $u$, due to the presence of the gravitational wave.

\subsection{An intermediate summary}
\label{sec-sum}

From the calculation in the previous subsections, one knows that if a standard Schwarzschild black hole is perturbed, its shear tensor is no longer zero.
In the case of the even parity perturbation, it is
\begin{subequations}
  \label{eq-shear-ss}
  \begin{equation}
    \label{eq-shear-e-ss}
    c_{AB}^{(e,n\ell)}(u)=e^{i\omega_{n\ell}u}Y_{AB}^{\ell 0},
  \end{equation}
  while for the odd parity perturbation,
  \begin{equation}
    \label{eq-shear-o-ss}
    c_{AB}^{(o,n\ell)}(u)=e^{i\omega_{n\ell}u}X_{AB}^{\ell 0},
  \end{equation}
\end{subequations}
where $\Psi$'s in Eqs.~\eqref{eq-shear-e-ss-0} and \eqref{eq-shear-o-ss-0} have been replaced by more concrete forms.
In addition, a particular overtone index $n$ is considered, so they are labeled by $n$, too.
Since the standard Schwarzschild black hole enjoys the exact spherical symmetry, $m$ is set to zero for simplicity.
The overall factors for these expressions are set to 1.
They can also be properly normalized, but this is not absolutely necessary for the following discussion.
So these two expressions serve as the basis for a general perturbation, given by,
\begin{equation}
  \label{eq-shear-g}
  c_{AB}=\sum_{n\ell}\left[A_{n\ell}c_{AB}^{(e,n\ell)}(u)+B_{n\ell}c_{AB}^{(o,n\ell)}(u)\right].
\end{equation}
The coefficients $A_{n\ell}$ and $B_{n\ell}$ shall be determined by initial conditions.
These basic tensors $c_{AB}^{(e,n\ell)}$ and $c_{AB}^{(o,n\ell)}$ are eigenfunctions of $\chi_t$, $\sD^2$ and $\chi_z$, with eigenvalues $i\omega_{n\ell}$, $4-\ell(\ell+1)$ and $im=0$, respectively.
They are also the eigenfunctions of $\sP$.
They will be called the standard bases for the future discussion.

Let $\gamma_A=(1,i\sin\theta)$ and $\bar\gamma_A=(1,-i\sin\theta)$.
The complex gravitational strain is \cite{Mitman:2020pbt}
\begin{equation}
  \label{eq-cstr}
  h=h_+-ih_\times=\frac{1}{2r}\bar\gamma^A\bar\gamma^Bc_{AB}.
\end{equation}
Since $Y^{\ell m}_{AB}$ and $X_{AB}^{\ell m}$ are related to the spin-weighted spherical harmonics ${}_{-2}Y_{\ell m}$ given by \cite{Nichols:2017rqr},
\begin{subequations}
  \begin{gather}
    Y_{AB}^{\ell m}=\frac{\mathsf N_\ell}{2}({}_{-2}Y_{\ell m}\gamma_A\gamma_B+{}_{2}Y_{\ell m}\bar\gamma_A\bar\gamma_B),\\
    X_{AB}^{\ell m}=\frac{\mathsf N_\ell}{i2}({}_{-2}Y_{\ell m}\gamma_A\gamma_B-{}_{2}Y_{\ell m}\bar\gamma_A\bar\gamma_B),
  \end{gather}
\end{subequations}
where $\mathsf N_\ell=\sqrt{(\ell-1)\ell(\ell+1)(\ell+2)}$ is a factor,
one has
\begin{equation}
  \label{eq-cstr-1}
  h=\frac{1}{4r}\sum_{n\ell}\mathsf N_le^{i\omega_{n\ell}u}(A_{n\ell}-iB_{n\ell})({}_{-2}Y_{\ell0}).
\end{equation}
With the matched filtering method, interferometers may detect the quasinormal modes of the black hole \cite{Berti:note}.
Many waveform templates implemented in LALSuite \cite{lalsuite} contain the coalescence phase $\phi_c$.
One can add its contribution to the above expression by multiplying it with $e^{i\phi_c}$, which can been absorbed by redefining $A_{n\ell}$ and $B_{n\ell}$.

\section{Supertranslated perturbations}
\label{sec-stper}

In the previous section, the black hole perturbation $p_{\mu\nu}$ have been written in the Bondi gauge, by solving the perturbed Einstein's equation.
However, the background metric is still the standard Schwarzschild metric $\mathring g_{\mu\nu}$.
In the current section, the whole metric $g_{\mu\nu}=\mathring g_{\mu\nu}+p_{\mu\nu}$ will be supertranslated to get $\hat g_{\mu\nu}$ with the general method in Sec.~\ref{sec-f-st}.
The background metric shall be supertranslated at the same time, resulting in $\hat{\mathring g}_{\mu\nu}$ as discussed in Sec.~\ref{sec-st-sbh}.
Then, the supertranslated perturbation is $\hat p_{\mu\nu}\equiv\hat g_{\mu\nu}-\hat{\mathring g}_{\mu\nu}$.
Following the method reviewed in Sec.~\ref{sec-f-st}, one easily finds out that the coordinate transformation rule for $g_{\mu\nu}\rightarrow\hat g_{\mu\nu}$ is similar to Eq.~\eqref{eq-st-sbh} at the leading orders in $1/r$.
The corrections due to the perturbation $p_{\mu\nu}$ occur at higher orders in $1/r$ in Eq.~\eqref{eq-st-sbh}.
So the coordinate transformation truncated at the orders explicit in Eq.~\eqref{eq-st-sbh} is enough to obtain $\hat g_{\mu\nu}=\hat{\mathring g}_{\mu\nu}+\hat p_{\mu\nu}$.
Since one is interested in quasinormal modes, one simply uses Eq.~\eqref{eq-c-st} to get the supertranslated shear tensor of $\hat g_{\mu\nu}$,
\begin{equation}
  \label{eq-st-shear}
  \begin{split}
    \hat c_{AB}=&\sum_{n\ell}\left[A_{n\ell}c_{AB}^{(e,n\ell)}(u+\beta)+B_{n\ell}c_{AB}^{(o,n\ell)}(u+\beta)\right]\\
    &-\sD_{\langle A}\sD_{B\rangle}\beta.
  \end{split}
\end{equation}
Here, the term in the second line appears in the supertranslated background metric $\hat{\mathring g}_{\mu\nu}$, Eq.~\eqref{eq-st-sbh-0}.
Therefore, the supertranslated tensor perturbation is given by the first line, and denoted as
\begin{equation}
  \label{eq-st-s-p}
  \delta\hat c_{AB}=\sum_{n\ell}\left[A_{n\ell}c_{AB}^{(e,n\ell)}(u+\beta)+B_{n\ell}c_{AB}^{(o,n\ell)}(u+\beta)\right],
\end{equation}
where
\begin{subequations}
  \label{eq-sh-st-0}
  \begin{gather}
    c_{AB}^{(e,n\ell)}(u+\beta)=\exp[i\omega_{n\ell}(u+\beta)]Y_{AB}^{\ell0},\label{eq-sh-st-e}\\
    c_{AB}^{(o,n\ell)}(u+\beta)=\exp[i\omega_{n\ell}(u+\beta)]X_{AB}^{\ell0}.\label{eq-sh-st-o}
  \end{gather}
\end{subequations}
Like $c_{AB}^{(e,n\ell)}(u)$ and $c_{AB}^{(o,n\ell)}(u)$ for the standard case, the above tensors serve as the bases for a generic perturbation about a supertranslated Schwarzschild black hole, so they will be named the supertranslated bases.

The supertranslated bases have the following interesting properties.
First, $c_{AB}^{(e,n\ell)}(u+\beta)$ and $c_{AB}^{(o,n\ell)}(u+\beta)$ are eigenfunctions of $\hat \chi_t$ with eigenvalues $i\omega_{n\ell}$, which can be verified directly.
Second, they are also eigenfunctions of the operators $\hat\sD^2$ and $\hat \chi_z$ with eigenvalues $4-\ell(\ell+1)$ and $im=0$, respectively.
Again, please refer to Appendix~\ref{sec-aop} for the definitions of these operators.
Third, they satisfy the correct boundary condition at the infinity.
Here, we only present the supertranslated perturbation near the infinity, so one shall check whether the complete perturbation satisfies the boundary condition near the horizon.
This is done in Appendix~\ref{app-bdyc}, where certain covariant boundary conditions, equivalent to Eq.~\eqref{eq-bdyc-qnm}, are proposed.
Finally, they have well-defined parities under $\hsP$, defined in Eq.~\eqref{eq-def-p-st}.
Indeed, under $\hsP$, one finds out that
\begin{equation*}
  \label{eq-hsp-ce}
  \begin{split}
    &\hsP_*c_{AB}^{(e,n\ell)}(u+\beta)\\
    =&\exp[i\omega_{n\ell}(\hsP u+\beta(\hsP\bm\theta))]Y_{AB}^{\ell 0}(\hsP\bm\theta)\\
    \approx&\exp\{i\omega_{n\ell}[u+\beta(\bm\theta)-\beta(\sP\bm\theta)+\beta(\sP\bm\theta)]\}(-1)^\ell Y_{AB}^{\ell0}(\bm\theta)\\
    =&(-1)^\ell c_{AB}^{(e,n\ell)}(u+\beta),
  \end{split}
\end{equation*}
where $\hsP_*$ is the pullback associated with $\hsP$ \cite{Wald:1984rg}.
Indeed, $c_{AB}^{(e,n\ell)}(u+\beta)$ is even with respect to $\hsP$.
Similarly, $c_{AB}^{(o,n\ell)}(u+\beta)$ is odd, i.e., $\hsP_*c_{AB}^{(o,n\ell)}(u+\beta)=(-1)^{\ell+1}c_{AB}^{(o,n\ell)}(u+\beta)$.
These properties also imply that the isospectrality still holds for the supertranslated Schwarzschild black hole, even though it is not explicitly spherically symmetric.
Therefore, the supertranslated bases share many similarities with the standard ones.

Unlike the standard perturbation bases, the supertranslated ones have the more complicated angular dependence.
This is because
the Killing vectors of the spatial rotations have different coordinate components for the standard and supertranslated black holes.
In particular, the coordinate system in Eq.~\eqref{eq-st-sbh-0} is not compatible with $\hat\chi_x$, $\hat\chi_y$ and $\hat\chi_z$.
So none of $c_{AB}^{(e,n\ell)}(u+\beta)$ and $c_{AB}^{(o,n\ell)}(u+\beta)$ is related to angular coordinates in simpler ways.

Due to the simple $u$-dependence of the bases, one can rewrite Eq.~\eqref{eq-sh-st-0} as
\begin{subequations}
  \begin{gather}
    c_{AB}^{(e,n\ell)}(u+\beta)=\exp[i\omega_{n\ell}\beta(\bm\theta)]c_{AB}^{(e,n\ell)}(u),\\
    c_{AB}^{(o,n\ell)}(u+\beta)=\exp[i\omega_{n\ell}\beta(\bm\theta)]c_{AB}^{(o,n\ell)}(u).
  \end{gather}
\end{subequations}
Thus, this transformation can also be viewed as the local phase rotation, similar to the $U(1)$ rotation in quantum electrodynamics \cite{Schwartz:2013pla}.
$\omega_{n\ell}$ plays the role of the $U(1)$ charge.
So the pair $n\ell$ and the parity $e/o$ label ``particle species" of the quasinormal modes.
Of course, for quasinormal modes, $\omega_{n\ell}$ is complex.
This kind of phase rotation is independent of the parity types.
We will not further pursue the analogy with the $U(1)$ gauge theory in the current work.

Now, compare Eq.~\eqref{eq-st-s-p} with Eq.~\eqref{eq-shear-g}, then it is easy to conclude that in order to get the supertranslated black hole perturbation, one simply shifts the retarded time from $u$ to $u+\beta(\bm\theta)$, irrelevant to the parity.
Since $\beta$ is a function of angular coordinates $\bm\theta$, this time shift is not uniform in the sky surrounding the black hole.
At different angular directions, the time shift is usually different.
The complex gravitational wave strain is now given by
\begin{equation}
  \label{eq-cst-st}
  \hat h=\frac{1}{4r}\sum_{n\ell}\mathsf N_\ell\exp\{i\omega_{n\ell}[u+\beta(\bm\theta)]\}(A_{n\ell}-iB_{n\ell})({}_{-2}Y_{\ell0}).
\end{equation}
Therefore, for different modes labeled by $n\ell$, the phase shifts are different, $\Re\omega_{n\ell}\beta(\bm\theta)$, with $\Re$ taking the real part.
For each fixed $\ell$, the phase shift increases with $n$.
Now, consider the decay of the supertranslated perturbation.
Each mode $n\ell$ decays according to $\exp(-\Im\omega_{n\ell}u)\exp[-\Im\omega_{n\ell}\beta(\bm\theta)]$ with $\Im$ extracting the imaginary part.
Here, the first exponential represents the standard decaying behavior, while the second is due to the supertranslation.
So compared with the standard black hole, the amplitude of the supertranslated perturbation is smaller if $\beta(\bm\theta)>0$, and larger if $\beta(\bm\theta)<0$, for a particular $u$.
As illustrated in the previous paragraph, the phase shift and the change in the amplitude are also angle dependent.

\subsection{Spherical decomposition}
\label{sec-sph-dec}

As discussed above, $c_{AB}^{(e,n\ell)}(u+\beta)$ and $c_{AB}^{(o,n\ell)}(u+\beta)$ form a basis for the supertranslated perturbation, and have definitive parities with respect to $\hsP$.
However, for an observer at the null infinity, it might be more natural to still use $\sP$ as the parity operator.
Then, none of $c_{AB}^{(e,n\ell)}(u+\beta)$ and $c_{AB}^{(o,n\ell)}(u+\beta)$ has a definitive parity, formally under the transformation $\sP$.
This is because $\beta(\bm\theta)$ appears in the exponents according to Eqs.~\eqref{eq-sh-st-e} and \eqref{eq-sh-st-o}.
To correctly determine the even and odd parity parts for $\sP$, one shall first perform the following expansions,
\begin{equation}
  \label{eq-spe-psi}
  \exp[i\omega_{n\ell}\beta(\bm\theta)]=\sum_{\breve\ell\breve m}\Phi^{n\ell}_{\beta|\breve\ell\breve m}Y^{\breve\ell\breve m},
\end{equation}
where $\Phi^{n\ell}_{\beta|\breve\ell\breve m}=\int\ud^2\theta\sqrt\gamma\exp[i\omega_{n\ell}\beta]\bar Y^{\breve\ell \breve m}$ are expansion coefficients.
In principle, $\breve \ell$ starts at $2$.
The cases of $\breve\ell=0$ correspond to the ordinary translations, not the proper supertranslations.
Moreover, these coefficients also depend on $\beta$, so $\Phi$'s are labeled by $\beta$, too.
After some mathematical manipulation, one finds out that
\begin{subequations}
  \label{eq-st-sh-eo-n}
  \begin{gather}
    c_{AB}^{(e,n\ell)}(u+\beta)=e^{i\omega_{n\ell}u}\sum_{\bar\ell\bar m}(\Lambda^{n\ell}_{\beta|\bar\ell\bar m}Y_{AB}^{\bar\ell\bar m}+\Xi^{n\ell}_{\beta|\bar\ell\bar m}X_{AB}^{\bar\ell\bar m}),\\
    c_{AB}^{(o,n\ell)}(u+\beta)=e^{i\omega_{n\ell}u}\sum_{\bar\ell\bar m}(\Lambda^{n\ell}_{\beta|\bar\ell\bar m}X_{AB}^{\bar\ell\bar m}-\Xi^{n\ell}_{\beta|\bar\ell\bar m}Y_{AB}^{\bar\ell\bar m}),
  \end{gather}
  where
  \begin{gather}
    \begin{split}
      \label{eq-def-lam}
      \Lambda^{n\ell}_{\beta|\bar\ell\bar m}=&\mathsf N_\ell\mathsf N_{\bar\ell}\sum_{\breve\ell}\left[1+(-1)^{\bar\ell+\ell+\breve\ell}\right]\Phi^{n\ell}_{\beta|\breve\ell\bar m}\\
      &\times \mathcal I_{\bar\ell}(0,\breve\ell,\bar m;-2,\ell,0),
    \end{split}\\
    \begin{split}
      \label{eq-def-xi}
      \Xi^{n\ell}_{\beta|\bar\ell\bar m}=&i\mathsf N_\ell\mathsf N_{\bar\ell}\sum_{\breve\ell}\left[1-(-1)^{\bar\ell+\ell+\breve\ell}\right]\Phi^{n\ell}_{\beta|\breve\ell\bar m}\\
      &\times \mathcal I_{\bar\ell}(0,\breve\ell,\bar m;-2,\ell,0).
    \end{split}
  \end{gather}
\end{subequations}
In these expressions, one defines \cite{Nichols:2017rqr}
\begin{equation}
  \label{eq-def-cl}
  \begin{split}
    &\mathcal I_{\bar\ell}(\breve s,\breve\ell,\breve m;s,\ell,m)\\
    \equiv&\int\ud^2\theta\sqrt\gamma({}_{\breve s}Y_{\breve\ell\breve m})({}_{s}Y_{\ell m})({}_{\bar s}\bar Y_{\bar\ell\bar m})\\
    =&(-1)^{\bar\ell+\breve\ell+\ell}\sqrt{\frac{(2\breve\ell+1)(2\ell+1)}{4\pi(2\bar\ell+1)}}\langle\breve\ell\breve s;\ell s|\bar\ell,\breve s+s\rangle\\
    &\times\langle\breve \ell\breve m;\ell m|\bar\ell,\breve m+m\rangle,
  \end{split}
\end{equation}
where the Clebsch-Gordon coefficients are used \cite{Cohen-Tannoudji:101367}.
This symbol vanishes unless $\bar s=\breve s+s$, $\bar m=\breve m+m$, and $\text{max}\{|\breve\ell-\ell|,|\breve m+m|,|\breve s+s|\}\le\bar\ell\le \breve\ell+\ell$.
It satisfies
\begin{equation}
  \label{eq-prop-cc}
  \begin{split}
    &\mathcal I_{\bar\ell}(\breve s,\breve\ell,\breve m;s,\ell,m)\\
    =&(-1)^{\bar\ell+\breve\ell+\ell}\mathcal I_{\bar\ell}(-\breve s,\breve\ell,\breve m_1;-s,\ell,m).
  \end{split}
\end{equation}
Since $\mathcal I_{\bar\ell}(0,\breve\ell,\bar m;-2,\ell,0)\propto \langle\breve\ell0;\ell,-2|\bar\ell,-2\rangle\langle\breve\ell\bar m;\ell0|\bar\ell\bar m\rangle$, the summations in Eqs.~\eqref{eq-def-lam} and \eqref{eq-def-xi} are from $|\bar\ell-\ell|$ to $\bar\ell+\ell$.

Therefore, by Eq.~\eqref{eq-st-sh-eo-n}, the supertranslated basic tensors are linear combinations of the following even and odd parity tensors,
\begin{subequations}
  \label{eq-nbs}
  \begin{gather}
    \hat c_{AB}^{(e,n\ell\bar\ell\bar m)}(u)=e^{i\omega_{n\ell}u}Y_{AB}^{\bar\ell\bar m},\label{eq-n-b-e}\\
    \hat c_{AB}^{(o,n\ell\bar\ell\bar m)}(u)=e^{i\omega_{n\ell}u}X_{AB}^{\bar\ell\bar m}.\label{eq-n-b-o}
  \end{gather}
\end{subequations}
These tensors are formally more similar to Eq.~\eqref{eq-shear-ss} than $c_{AB}^{(e,n\ell)}(u+\beta)$ and $c_{AB}^{(o,n\ell)}(u+\beta)$.
These are also eigenfunctions of $\hat\chi_t$ with eigenvalues $i\omega_{n\ell}$, so these basic tensors are also oscillating at $\Re\omega_{n\ell}$ and decaying at the rates $\Im\omega_{n\ell}$.
Unfortunately, they fail to be the eigenfunctions of $\hat\sD^2$ and $\hat\chi_z$ of the exact SO(3) symmetry of the supertranslated Schwarzschild metric.
Therefore, one does not use them to decompose $\delta c_{AB}$, as usual.
However, as argued previously, it is always convenient to decompose an arbitrary symmetric, traceless tensor (e.g. $\delta\hat c_{AB}$) in terms of such tensors.
Moreover, it is a common practice to reexpress tensors on a celestial sphere in terms of spherical harmonics, even if the spacetime is not exactly spherically symmetric \cite{Maggiore:1900zz,Poisson2014}.
Let us call the tensors in Eq.~\eqref{eq-nbs} the pseudo basic tensors.

So if one insists on making the decomposition~\eqref{eq-st-sh-eo-n}, one knows that  $c_{AB}^{(e,n\ell)}(u+\beta)$ and $c_{AB}^{(o,n\ell)}(u+\beta)$ are linear combinations of the even and odd parity pseudo basic tensors.
Both the even and odd parity pseudo basic tensors  have the same spectrum of the complex frequencies $\omega_{n\ell}$, so the isospectrality still holds for the supertranslated black hole.
Moreover, the pseudo basic tensors are still the eigenfunctions of the operators $\sD^2$ and $\chi_z$ with the eigenvalues listed in the last column of Table~\ref{tab-ev-ef}.
Of course, the pseudo bases carry more indices, and the pair $n\ell$ is not related to $\bar\ell\bar m$.
This reflects the fact that they do not serve as a set of good bases for the solutions to the perturbed Einstein's equation about the supertranslated Schwarzschild metric.

\subsubsection{Special cases with the cylindrical symmetry}
\label{sec-s-azi}

The spherical decomposition of a generic quasinormal mode for the supertranslated metric with the cylindrical symmetry discussed in Sec.~\ref{sec-azi}  can also be done as in the previous subsection.
To simplify the discussion, let us consider $\beta=\beta(\theta)$ without the loss of generality.
In this case, $\Phi$'s introduced in Eq.~\eqref{eq-spe-psi} are independent of $\breve m$, so one can set $\breve m=0$.
And $\bar m$'s in Eq.~\eqref{eq-st-sh-eo-n} shall be also be set to 0.
Therefore, the supertranslated bases can be linearly expanded with the following pseudo basic tensors,
\begin{subequations}
  \begin{gather}
    \hat c_{AB}^{(e,n\ell\bar\ell)}=e^{i\omega_{n\ell}u}Y_{AB}^{\ell0},\\
    \hat c_{AB}^{(o,n\ell\bar\ell)}=e^{i\omega_{n\ell}u}X_{AB}^{\ell0}.
  \end{gather}
\end{subequations}
With respect to these bases, one may still claim that the even and odd parity basic perturbations share the same set of the complex frequencies $\omega_{n\ell}$.
Again, as in the previous subsection, these pseudo basic tensors are convenient, but they cannot be serve as the good bases for the solutions to the perturbed Einstein's equation with the azimuthal symmetry.

\section{Speculation on detection}
\label{sec-det1}

The detection of the quasinormal mode of the supertranslated Schwarzschild black hole is based on the complex gravitational wave strain, Eq.~\eqref{eq-cst-st}, which is rewritten in the following way,
\begin{equation}
  \label{eq-cst-st-1}
  \begin{split}
    \hat h=&\frac{1}{4r}\sum_{n\ell}\mathsf N_\ell(A_{n\ell}-iB_{n\ell})({}_{-2}Y_{\ell0})\\
    &\times\exp\left[-\frac{u+\beta(\bm\theta)}{\tau_{n\ell}}\right]\exp\left\{i\varpi_{n\ell}\big[u+\beta(\bm\theta)\big]\right\},
  \end{split}
\end{equation}
where $\tau_{n\ell}=1/\Im\omega_{n\ell}$ and $\varpi_{n\ell}=\Re\omega_{n\ell}$.
As discussed previously, the supertranslation changes the amplitude of the quasinormal mode $n\ell$  by $\exp[-\beta(\bm\theta)/\tau_{n\ell}]$ and causes the phase shift $\varpi_{n\ell}\beta(\bm\theta)$, anisotropically.
So for each mode $n\ell$, although the decay rate does not change, the profile of the amplitude as a function of the time $u$ is shifted by $\beta(\bm\theta)$.
Since the dominant mode with $n=0$ and $\ell=2$ has the longest $\tau_{02}$ \cite{Chandrasekhar:1985kt}, the change in its amplitude is the least.
This mode oscillates the most frequently, so its phase shift is the largest.

It is possible that the interferometer detects the supertranslated quasinormal mode, and measures $\beta(\bm\theta)$ with the matched filtering method.
This method relies on the construction of the waveform template.
For the supertranslated quasinormal mode, it is easy to modify the existing waveform template, that is, just shifting the time axis by $\beta(\bm\theta)$ in the time domain for the ringdown phase.
In the frequency domain, one shall shift the ringdown phase by $\varpi_{n\ell}\beta(\bm\theta)$, and multiply the amplitude by a factor of $\exp[-\beta(\bm\theta)/\tau_{n\ell}]$ for each quasinormal mode $n\ell$.

As identified before, the phase shift for the mode $n\ell$ is
\begin{equation}
  \label{eq-dep-nl}
  \delta\phi_{n\ell}=\varpi_{n\ell}\beta(\bm\theta).
\end{equation}
Since interferometers are very sensitive to phase shift, it would be easier to measure it to infer $\beta(\bm\theta)$.
Since the supertranslation considered here is finite, $\beta(\bm\theta)$ can be large, so the phase shift $\delta\phi_{n\ell}$ can also be large, and this might make the measurement easier.
Moreover, $\delta\phi_{n\ell}$ is not isotropic, which suggests to use several interferometers to detect the supertranslated quasinormal modes.
This is important, especially when the interferometers may be capable of observing only the dominant quasinormal mode $nl=02$ in the near future.
Because if there is only one interferometer that successfully measures $\delta\phi_{02}=\varpi_{02}\beta(\bm\theta_0)$, one cannot be so confident that it comes from the supertranslation.
It can be set to zero by absorbing it into various phases already implemented in many waveform templates \cite{Husa:2015iqa,Khan:2015jqa,Pratten:2020ceb}, or equivalently, shifting the time axis by $\beta(\bm\theta_0)$ \footnote{
  Note that since the time shift $\beta(\bm\theta_0)$ is a constant, it is an ordinary time translation, not a proper supertranslation.
  So it does not change the background metric or the supertranslated quasinormal mode.
}.
Instead, if there are multiple interferometers measuring several phase shifts $\delta\phi_{02}^{(k)}=\varpi_{02}\beta(\bm\theta_k)$ ($k=0,1,2,\cdots$), the difference between any pair of the phase shifts is intrinsic.
That is, even if one can shift the time $u$ such that one of the phase shifts, say $\delta\phi_{02}^{(0)}$, becomes zero, the differences
\begin{equation}
  \label{eq-dif}
  (\delta\phi_{02}^{(k_1)}-\delta\phi_{02}^{(0)})-(\delta\phi_{02}^{(k_2)}-\delta\phi_{02}^{(0)})=\delta\phi_{02}^{(k_1)}-\delta\phi_{02}^{(k_2)},
\end{equation}
are invariant.
We emphasize that one shall shift all time axes for these interferometers by the same amount, otherwise, the time shift would be angular dependent, $u\rightarrow u+\beta'(\bm\theta)$.
$\beta'(\bm\theta)$ may actually parameterize another supertranslation, and this changes the background spacetime metric, which is given by Eq.~\eqref{eq-st-sbh-0} with $\beta$ replaced by $\beta+\beta'$.
In particular, the shear tensor changes, $\hat{\mathring c}'_{AB}=\hat{\mathring c}_{AB}-2\sD_{\langle A}\sD_{B\rangle}\beta'$.
This is not the metric that one started with.
Since we are interested in the quasinormal modes of the original metric, it is not permissible to shift the time axes of the interferometers independently.
Of course, if one insists on shifting the time axes independently, one actually measures the quasinormal modes for a different supertranslated Schwarzschild black hole, and the phase shifts become $\delta\phi'^{(k)}_{02}=\delta\phi^{(k)}_{02}+\varpi_{02}\beta'(\bm\theta_k)$.
So
\begin{equation}
  \label{eq-st-pd}
  \begin{split}
    \delta\phi'^{(k_1)}_{02}-\delta\phi'^{(k_2)}_{02}=&\delta\phi^{(k_1)}_{02}-\delta\phi^{(k_2)}_{02}\\
    &+\varpi_{02}[\beta'(\bm\theta_{k_1})-\beta'(\bm\theta_{k_2})].
  \end{split}
\end{equation}
The second line does not vanish in general.
These phase differences depend on the supertranslation, and truly characterize the supertranslated spacetime.

Therefore, one shall use multiple interferometers in order to determine whether there exists a supertranslated black hole.
Of course, if the supertranslated perturbation is gravitationally lensed by a massive star or a galaxy \cite{gravlens1992}, it is possible to used a single interferometer to detect the supertranslated black hole, as the lensed gravitational wave would travel in multiple paths to reach the earth.
It is understood that, in order to measure the phase difference more easily, it is better to have a lensing system in which all objects are close enough, so that the angular separations between gravitational wave rays are sufficiently large.

\section{Conclusion}
\label{sec-con}

In this work, we considered the supertranslated black hole and its quasinormal modes.
Such kind of black holes are related to the standard Schwarzschild black holes via suitable (finite) supertranslations.
Since a supertranslation usually modifies the asymptotic charges, i.e., the super-Lorentz charges, it is believed that the supertranslated black hole is physically distinguishable from the standard one.
Since a generic gravitational collapsing would result in the supertranslated black hole, it is interesting to study whether there is some method to detect its existence.
Given that the gravitational wave has been detected \cite{gw150914,gw170817}, the quasinormal modes of the supertranslated black hole were discussed.

We started with the supertranslated Schwarzschild black hole, due to its simplicity.
Although Ref.~\cite{Compere:2016hzt} already derived the closed form metric for such a black hole, we did not work out its perturbation equation due to its extreme complexity and its lack of the apparent spherical symmetry.
Instead, we made use of the general covariance of Einstein's equation, and obtained the quasinormal modes by suitably supertranslating the known perturbations of the standard Schwarzschild black hole.
This method can be easily generalized to the Kerr black hole.

We thus identified the shear tensor associated with the supertranslated perturbation, which turns out to be given by shifting the standard result by $\beta(\bm\theta)$ in time.
The supertranslated tensor basis can also be obtained in the same way.
These bases are also time translation invariant, eigenfunctions of rotation operators and a suitably defined asymptotic parity operator $\hsP$.
These bases share the same spectra of the complex frequencies, and the isospectrality also holds.
We also considered the pseudo tensor basis, which is convenient, especially when one does not know for sure if the background spacetime is described by a supertranslated metric.
With respect to the pseudo tensor basis, the isospectrality also holds formally.

With the gravitational wave interferometers, one may measure the phase shift proportional to $\beta(\bm\theta)$ of the gravitational wave strain during the ringdown stage.
Due to the freedom of performing the time translation, one shall use multiple detectors, and if the differences in the phase shifts are nonvanishing, one may infer the possible existence of the supertranslated black hole.
A single detector may also achieve this goal if the supertranslated quasinormal mode is gravitationally lensed.

\begin{acknowledgements}
  This work was supported by the National Natural Science Foundation of China under Grant No.~11633001 and No.~11920101003, the Key Program of the  National Natural Science Foundation of China under Grant No.~12433001, and the Strategic Priority Research Program of the Chinese Academy of Sciences, Grant No.~XDB23000000.
  S. H. was supported by the National Natural Science Foundation of China under Grant No.~12205222.
  K.L. was supported by Fapesq-PB of Brazil.
\end{acknowledgements}

\appendix

\section{Generalizing the angular momentum operators}
\label{sec-aop}

In the treatment of the black hole perturbations of the standard Schwarzschild black hole, the perturbations are conveniently written as the linear combinations of eigenfunctions of operators $\pd_u$, $\sD^2$, $\chi_i$'s, and $\sP$ due to the nice symmetries possessed by the background metric.
For the supertranslated black hole, the metric is not explicitly invariant under the spatial rotations.
As discussed previously, the supertranslated metric is still implicitly invariant under the transformations generated by $\hat\chi_i$'s.
Although the perturbation to the supertranslated black hole may not be decomposed into the eigenfunctions of $\sD^2$ and $\chi_i$'s, it is still possible to find the generalizations of these operators such that the perturbation is the linear combination of their eigenfunctions.

The generalization is based on the observation that for the standard Schwarzschild metric, operators $\chi_i$'s can be viewed as Lie derivatives,
\begin{equation}
  \label{eq-op-lie}
  \chi_iT_{\cdots}:=\lie_{\chi_i}T_{\cdots},
\end{equation}
when they act on a tensor field $T_{\cdots}$ defined on a unit 2-sphere.
The indices of $T$ are represented by $\cdots$.
This definitely works for a scalar field $T$,
\begin{equation}
  \label{eq-cs-g}
  \chi_iT=\chi_i^A\pd_AT=\lie_{\chi_i}T.
\end{equation}
For the higher rank tensors $T_{\cdots}$, one takes Eq.~\eqref{eq-op-lie} as the definition of the angular momentum operators.
The Laplacian operator $\sD^2$ can be reexpressed in terms of the Lie derivatives.
For a scalar field $T$, it is simply
\begin{equation}
  \label{eq-def-lap-0}
  \sD^2T=\sum_i\lie_{\chi_i}\lie_{\chi_i}T.
\end{equation}
For a rank-2 tensor $T_{AB}$, one has
\begin{equation}
  \label{eq-def-lap-2}
  \sD^2T_{AB}=\sum_i\lie_{\chi_i}\lie_{\chi_i}T_{AB}+4T_{\langle AB\rangle}-2\epsilon_{AB}\epsilon^{CD}T_{CD}.
\end{equation}
One can check that these definitions agrees with the familiar actions of the angular momentum operators on a rank-2 symmetric, traceless tensors $T_{AB}$.
Such a $T_{AB}$ can be written as a linear combination of $Y_{AB}^{\ell m}$ and $X_{AB}^{\ell m}$.
It can be shown directly that
\begin{subequations}
  \label{eq-lcs}
  \begin{gather}
    \lie_{\chi_i}Y_{AB}^{\ell m}=\sD_{\langle A}\sD_{B\rangle}[\chi_i^C\pd_CY^{\ell m}],\label{eq-lcy}\\
    \lie_{\chi_i}X_{AB}^{\ell m}=-\epsilon_{(A}{}^C\sD_{B)}\sD_C[\chi_i^C\pd_CY^{\ell m}].\label{eq-lcx}
  \end{gather}
\end{subequations}
Therefore, for $\chi_z$, one has
\begin{subequations}
  \begin{gather}
    \lie_{\chi_z}Y_{AB}^{\ell m}=-imY_{AB}^{\ell m},\\
    \lie_{\chi_z}X_{AB}^{\ell m}=-imX_{AB}^{\ell m}.
  \end{gather}
\end{subequations}
Indeed, $Y_{AB}^{\ell m}$ and $X_{AB}^{\ell m}$ are eigenfunctions of $\lie_{\chi_z}$.
Now, one uses Eq.~\eqref{eq-def-lap-2} to check
\begin{equation}
  \begin{split}
    \sD^2Y_{AB}^{\ell m}=&\sum_i\sD_{\langle A}\sD_{B\rangle}[\chi_i^D\sD_D(\chi_i^C\sD_CY^{\ell m})]+4Y_{AB}^{\ell m}\\
    =&[4-\ell(\ell+1)]Y_{AB}^{\ell m},
  \end{split}
\end{equation}
and a similar equation for $X_{AB}^{\ell m}$ can also be obtained.
Here, one shall repeatedly use Eqs.~\eqref{eq-lcy} and \eqref{eq-def-lap-0}.
Comparing the above result with Table~\ref{tab-ev-ef}, one knows that the definition \eqref{eq-def-lap-2} works.

For the supertranslated black hole, one shall generalize $\chi_i$  in the following way,
\begin{equation}
  \label{eq-ge-anp}
  \chi_iT_{\cdots}\mapsto \hat\chi_iT_{\cdots}:=\fL_{\hat\chi_i}T_{\cdots}.
\end{equation}
Instead of simply using the Lie derivatives $\lie_{\hat\chi_i}$'s, the actions of the BMS generators $\hat\chi_i$'s are used.
So for a scalar field $T$, one has \cite{Hou:2020tnd,Hou:2021oxe}
\begin{equation}
  \label{eq-def-bms-s0}
  \fL_{\hat\chi_i}T=-(\chi_i^A\pd_A\beta)\dot T+\lie_{\chi_i}T.
\end{equation}
For the symmetric, traceless tensor $T_{AB}$, one defines
\begin{equation}
  \label{eq-def-bms-stf}
  \fL_{\hat\chi_i}T_{AB}=-(\chi_i^A\pd_A\beta)\dot T_{AB}+\lie_{\chi_i}T_{AB},
\end{equation}
following the transformation of $c_{AB}$, Eq.~\eqref{eq-btf-c}.
Note that for a rotation generator, $\psi=0$.
When $\beta\rightarrow0$, Eqs.~\eqref{eq-def-bms-s0} and \eqref{eq-def-bms-stf} reduce to Eq.~\eqref{eq-op-lie}.
With $\fL_{\hat\chi_i}$, one defines the Laplacian operator $\hat\sD^2$ in the supertranslated case as,
\begin{equation}
  \label{eq-def-slap}
  \hat\sD^2T=\sum_i\fL_{\hat\chi_i}\fL_{\hat\chi_i}T,
\end{equation}
for a scalar $T$, and
\begin{equation}
  \label{eq-def-tlap}
  \hat\sD^2T_{AB}=\sum_i\fL_{\hat\chi_i}\fL_{\hat\chi_i}T_{AB}+4T_{AB},
\end{equation}
for a symmetric, traceless tensor $T_{AB}$.

In Section~\ref{sec-stper}, the supertranslated bases \eqref{eq-sh-st-0} were introduced.
Here, one can show that these are eigenfunctions of $\fL_{\hat\chi_z}$ and $\hat\sD^2$.
Indeed, one can find out that
\begin{subequations}
  \begin{gather}
    \fL_{\hat\chi_z}c_{AB}^{(e,n\ell)}(u+\beta)=0,\\
    \fL_{\hat\chi_z}c_{AB}^{(o,n\ell)}(u+\beta)=0,
  \end{gather}
  and
  \begin{gather}
    \hat\sD^2c_{AB}^{(e,n\ell)}=[4-\ell(\ell+1)]c_{AB}^{(e,n\ell)},\\
    \hat\sD^2c_{AB}^{(o,n\ell)}=[4-\ell(\ell+1)]c_{AB}^{(e,n\ell)}.
  \end{gather}
\end{subequations}
To obtain these results, one shall use Eq.~\eqref{eq-lcs}.

\section{The covariant boundary conditions}
\label{app-bdyc}

Equation~\eqref{eq-sh-st-0} shows the supertranslated bases for quasinormal modes.
One concludes that both $c^{(e,n\ell)}_{AB}(u+\beta)$ and $c^{(o,n\ell)}_{AB}(u+\beta)$ are outgoing, satisfying the boundary condition of the quasinormal mode at the infinity.
In this appendix, we will show that the boundary conditions at the horizon and the infinity are both satisfied after an arbitrary coordinate transformation.

For that purpose, one shall reexpress the boundary condition in a covariant manner.
The boundary condition given by \eqref{eq-bdyc-qnm} actually states that at both the horizon and the infinity, the gravitational energy flux shall propagate into the horizon and the infinity.
Given the phase of the gravitational wave $\Phi$, the direction of the flux can be represented by
\begin{equation}
  \label{eq-def-k}
  k^\mu=-\nabla^\mu\Phi,
\end{equation}
near the horizon and the infinity, according to the short-wavelength approximation \cite{mtw,Isaacson:1967zz}.
Note that near the horizon, the wave is highly blueshifted relative to the local inertia observer, and near the infinity, the background spacetime is basically Minkowski.
So the short-wavelength approximation is satisfied.

Next, let us define a level function $\Omega$, taking the advantage of the spherical symmetry of the Schwarzschild spacetime.
With this level function $\Omega$, the spacetime is foliated into leaves, which are sets of the orbits of the isometry group SO(3) \cite{Wald:1984rg}.
The horizon occurs at $\Omega_H$, and as one moves away from the horizon to the infinity, $\Omega$ increases monotonically and becomes infinite at the infinity.
Note that the Schwarzschild spacetime is always spherically symmetry even if one does not use the (standard) Schwarzschild coordinates to describe it.
In a coordinate system adapted to the symmetry, the metric is explicitly spherically symmetric, while in a coordinate system not compatible to the symmetry, the metric is implicitly spherically symmetric.
So in any coordinate system, one can always define such a level function $\Omega$ and foliate the spacetime accordingly.

Now, with the wave vector $k^\mu$ and the level function $\Omega$, one can write down the boundary condition at the horizon
\begin{equation}
  \label{eq-bdy-h}
  \lim_{\Omega\rightarrow\Omega_H}k^\mu\nabla_\mu\Omega<0.
\end{equation}
Near the infinity, one has
\begin{equation}
  \label{eq-bdy-inf}
  \lim_{\Omega\rightarrow\infty}k^\mu\nabla_\mu\Omega>0,
\end{equation}
instead.

One can show that Eq.~\eqref{eq-bdyc-qnm} agrees with these covariant boundary conditions.
In Eq.~\eqref{eq-bdyc-qnm}, the adapted coordinate system is used, so the level function can be taken to be $\Omega=r$.
Near the infinity, the phase of the wave is $\Phi=\varpi u\approx\varpi(t-r)$ with $\varpi$ the real frequency, so the wave vector is $k^\mu\approx-\varpi (\delta^\mu_t-\delta^\mu_r)$, and $\lim_{\Omega\rightarrow\infty}k^\mu\nabla_\mu\phi=\varpi>0$.
Near the horizon, the phase is given by $\Phi=\varpi v=\varpi(t+r_*)$, so one has $k^\mu=-\varpi(g^{\mu t}+g^{\mu r}\ud r_*/\ud r)$, and $\lim_{\Omega\rightarrow\Omega_H}k^\mu\nabla\Omega=\lim_{\Omega\rightarrow\Omega_H}-\varpi g^{rr}\ud r_*/\ud r=-\varpi<0$.

On the other hand, one can use the covariant boundary conditions to reproduce Eq.~\eqref{eq-bdyc-qnm}.
Indeed, for any phase $\Phi$, one has $\nabla_\mu\Phi=\Phi_t\nabla_\mu t+\Phi_r\nabla_\mu r$.
It has no components in the angular directions because the angular dependence is encoded in the tensor spherical harmonics.
Near the infinity, $\lim_{\Omega\rightarrow\infty}k^\mu\nabla_\mu\Omega=-\Phi_r|_\infty>0$, and the null condition $k^\mu k_\mu=0$ implies that $\Phi_t|_\infty=-\Phi_r|_\infty$.
The negative sign is due to the fact that $k^\mu$ shall be future-pointing.
So $\nabla_\mu\Phi|_\infty=\Phi_t|_\infty\nabla_\mu(t-r)=\Phi_t|_\infty\nabla_\mu u$, and $\Phi|_\infty=\Phi_t|_\infty u$, modulo a constant phase.
Near the horizon, $\lim_{\Omega\rightarrow\Omega_H}k^\mu\nabla_\mu\Omega=-\lim_{r\rightarrow r_H}\Phi_rg^{rr}<0$, and $\Phi_t=\Phi_rg^{rr}$ due to the null condition.
So, $\nabla_\mu\Phi=\Phi_rg^{rr}(\nabla_\mu t+g_{rr}\nabla_\mu r)=\Phi_rg^{rr}(\nabla_\mu t+\nabla_\mu r_*)=\Phi_rg^{rr}\nabla_\mu v$, which implies that $\lim_{\Omega\rightarrow\Omega_H}\Phi=\lim_{r\rightarrow r_H}(\Phi_rg^{rr})v$.

In the end, one may generalize the above covariant boundary conditions to more general level functions $\Omega'$, as long as
\begin{equation*}
  \label{eq-def-op}
  \Omega'=\left\{
  \begin{array}{cc}
    \Omega+(\Omega-\Omega_H)a+O(\Omega-\Omega_H)^2 & , \Omega\rightarrow\Omega_H, \\
    \Omega+O(1/\Omega)                             & , \Omega\rightarrow\infty,
  \end{array}
  \right.
\end{equation*}
where $a>-1$ is independent of $\Omega$.

\bibliography{qnmSupSch_v5.bbl}

\end{document}